\documentclass[11pt, a4paper, logo]{googledeepmind}

\usepackage[
    natbib=true,
    backend=biber,
    style=numeric,
]{biblatex}
\AtEveryBibitem{\clearfield{month}}
\AtEveryBibitem{\clearfield{day}}
\usepackage{csquotes}

\title{\emph{De novo} design of high-affinity protein binders with AlphaProteo}

\correspondingauthor{
\hspace{-1.5mm} send correspondence regarding this report to alphaproteo@google.com.}

\renewcommand{\today}{2024-09-05}

\author[*,1]{Vinicius Zambaldi}
\author[*,1]{David La}
\author[*,1]{Alexander E. Chu}
\author[*,1]{Harshnira Patani}
\author[*,1]{Amy E. Danson}
\author[*,1]{Tristan O. C. Kwan}
\author[*,1]{Thomas Frerix}
\author[*,1]{Rosalia G. Schneider}
\author[*,1]{David Saxton}
\author[*,1]{Ashok Thillaisundaram}
\author[*,1]{Zachary Wu}
\author[2]{Isabel Moraes}
\author[2]{Oskar Lange}
\author[1]{Eliseo Papa}
\author[1]{Gabriella Stanton}
\author[1]{Victor Martin}
\author[1]{Sukhdeep Singh}
\author[1]{Lai H. Wong}
\author[2]{Russ Bates}
\author[2]{Simon A. Kohl}
\author[1]{Josh Abramson}
\author[1]{Andrew W. Senior}
\author[3]{Yilmaz Alguel}
\author[4]{Mary Y. Wu}
\author[5]{Irene M.\,Aspalter}
\author[5,6]{Katie Bentley}
\author[7]{David L.V. Bauer}
\author[3]{Peter Cherepanov}
\author[1]{Demis Hassabis}
\author[1]{Pushmeet Kohli}
\author[1,$\dag$]{Rob Fergus}
\author[1,$\dag$]{Jue Wang}

\affil[*]{Equal contributions}
\affil[$\dag$]{Equal supervision}
\affil[1]{Google DeepMind}
\affil[2]{Work performed while at Google DeepMind} 
\affil[3]{The Chromatin Structure and Mobile DNA Laboratory, The Francis Crick Institute, London, UK} 
\affil[4]{COVID Surveillance Unit, The Francis Crick Institute, London, UK} 
\affil[5]{Cellular Adaptive Behaviour Laboratory, The Francis Crick Institute, London, UK.} 
\affil[6]{Department of Informatics, King’s College London, London, UK. K.B. performed the work at the Cellular Adaptive Behaviour Laboratory, The Francis Crick Institute, London, UK}
\affil[7]{RNA Virus Replication Laboratory,
The Francis Crick Institute, London, UK} 

\usepackage{pdflscape}
\usepackage{xcolor}
\usepackage{textcomp}
\usepackage{rotating}
\usepackage{setspace}

\usepackage{graphicx}
\usepackage{subcaption}
\usepackage{caption}
\newcommand{\titledcaption}[2]{\caption[{#1}]{\textbf{#1}\\#2}}

\usepackage{longtable}
\usepackage{booktabs}
\usepackage{makecell}
\newcommand{\mc}[2]{\makecell[t]{#1 \\ \scriptsize{(#2)}}}
\newcommand{\mcb}[2]{\makecell[t]{{\bfseries #1} \\ \scriptsize{(#2)}}}
\newcommand{\nan}{\makecell{--}}
\usepackage{array}
\usepackage{threeparttable}
\usepackage{multirow}

\usepackage{amsmath}
\usepackage{siunitx}

\usepackage{enumitem}
\usepackage{float}
\usepackage{seqsplit}

\usepackage{hyperref}
\hypersetup{
  colorlinks=true,
  linkcolor=black,
  citecolor=black,
  filecolor=black,
  urlcolor=black
}

\usepackage{tocloft}

\usepackage{etoolbox}
\makeatletter
\patchcmd{\@tocline}
  {\hfil}
  {\leaders\hbox{\hfil}\hfil}
  {}{}
\makeatother

\AtBeginDocument{

}

\newcommand{\PE}[2]{\textrm{PE}_{\textrm{FITC{#1},{#2}target}}}
\newcommand{\EC}{$\text{EC}_{50}$}
\newcommand{\KD}{{$\mathrm{K_D}$}}
\newcommand{\Tm}{{$\mathrm{T_m}$}}

\begin{abstract}
Computational design of protein-binding proteins is a fundamental capability with broad utility in biomedical research and biotechnology.
Recent methods have made strides against some target proteins, but on-demand creation of high-affinity binders without multiple rounds of experimental testing remains an unsolved challenge.
This technical report introduces AlphaProteo, a family of machine learning models for protein design, and details its performance on the \emph{de novo} binder design problem. With AlphaProteo, we achieve 3- to 300-fold better binding affinities and higher experimental success rates than the best existing methods on seven target proteins.
Our results suggest that AlphaProteo can generate binders "ready-to-use" for many research applications using only one round of medium-throughput screening and no further optimization.
\end{abstract}

\begin{document}
\maketitle
\begin{refsection} 
\section*{Experimental highlights}
\begin{itemize}
    \item We introduce the AlphaProteo protein design system and experimentally test binders designed against eight structurally diverse target proteins.
    \item For seven of the targets, between 9\% and 88\% of the designs tested in the wet lab were experimentally verified as successful binders. These figures are higher than the best existing method and 5- to 100-fold higher than other methods. For one of these targets we report the first computationally designed binders.
    \item The \emph{in silico} performance of AlphaProteo on hundreds of target proteins from the PDB is comparable to these seven targets, suggesting that the method can potentially generalize widely. We chose one of the most challenging targets from this PDB screen as an 8th target but failed to obtain binders.
    \item We obtain binders with 80-960 picomolar affinities to four targets and low-nanomolar affinities to another three without needing high-throughput screening or experimental affinity optimization. For the seven targets, our designs have 3- to 300-fold better binding affinities than the best previous designed binder.
    \item We test binders for two of our targets for biological function, demonstrating inhibition of VEGF signaling in human cells and SARS-CoV-2 neutralisation in Vero monkey cells.
    \item Cryo-EM and X-ray crystallography confirm the designed binder and binder-target complex structures.
\end{itemize}

\clearpage

\tableofcontents
\enlargethispage{1cm} 
\thispagestyle{empty}

\clearpage

\section{Introduction}
\label{sec:introduction}
Protein-protein interaction is a fundamental aspect of protein function, and protein-binding proteins are a basic building block for therapeutics, diagnostics, and biomedical research~\citep{Janin2008-ji, Marchand2022-bb}.
Traditionally, antibodies, nanobodies, and other scaffolds such as DARPins are developed into binders against a wide range of targets by immunization or directed evolution~\citep{Qian2023-va, Muyldermans2021-ne, Gebauer2020-tf}.
However, experimental selection does not afford control over the target epitope and is often too laborious for routine research applications.
Computational design of binders \emph{de novo}, without using a natural protein as a starting point, can target pre-specified epitopes and generate binders that are smaller, more thermostable, and easier to express than antibodies~\citep{Fleishman2011-vo, Silva2019-ir, Cao2022-vw}.

Recently, deep-learning based models have achieved major advances in biomolecular structure prediction~\citep{Jumper2021-sa, Baek2021-tc, Lin2023-rr, Krishna2024-xw, Abramson2024-nx} and protein design~\citep{Ingraham2023-yg, Watson2023-fy, Ruffolo2024-ax, Hayes2024-sp, Chu2024-md, Notin2024-ym}.
This has enabled progress on key scientific and societal challenges~\citep{Kovalevskiy2024-bh}, including the prediction and design of protein-protein interactions~\citep{Evans2021-ln, Humphreys2021-uh, Bryant2022-cg, Watson2023-fy, Gainza2023-nk, Goudy2023-ir, Dauparas2022-fc}.
It is now possible to obtain computationally designed binders to some targets without high-throughput screening~\citep{Watson2023-fy, Goudy2023-ir, Gainza2023-nk}.
High binding affinity without experimental optimization has also been achieved in some cases, such as for small peptides or disordered targets~\citep{Vazquez-Torres2024-uq, Wu2024-jo}.
However, success rates remain low against convex or polar epitopes, the affinity of the initial designs is usually poor, and many targets remain intractable~\citep{Yang2024-dq, Berger2024-wf}.

In this technical report focusing solely on experimental validation, we present the AlphaProteo protein design system and show that it can design \emph{de novo} protein-binding proteins with the following advantages:
\begin{enumerate}
    \item \textbf{High success rate}: stable, highly expressed, and specific binders can be obtained from screening tens of design candidates, alleviating the need for high-throughput methods.
    \item \textbf{High affinity}: for every target tested except one, the best binders have sub-nanomolar or low-nanomolar binding affinity (\KD), minimizing the labor needed for downstream affinity optimization.
    \item \textbf{General}: binders are successfully obtained against a range of targets with diverse structural and biochemical properties, using a single design method without complex manual intervention.
\end{enumerate}

\section{Results}
\label{sec:results}
AlphaProteo comprises two components (\autoref{fig:overview:pipeline_schematic}): a generative model trained on structure and sequence data from the Protein Data Bank (PDB) and a distillation set of AlphaFold predictions, as well as a filter which scores generated designs to predict whether they will succeed experimentally.
To design binders, we input a structure of the "target" protein and optionally designate "hotspot" residues representing the target epitope; the generative model outputs a structure and sequence of a candidate binder for that target (\autoref{fig:overview:binder_design_schematic}).
We generate a large number of design candidates and then filter them to a smaller set prior to experimental testing. The generative model compares favorably to the best existing method on \emph{in silico} benchmarks (\autoref{fig:in_silico_results}, \autoref{sec:in_silico_benchmark}).

\subsection{Sub-nanomolar-affinity binders from medium-throughput screening}
\begin{figure}[H]
    \centering
    \begin{subfigure}[t]{0.49\textwidth}
        \centering
        \caption*{\textbf{A}}
        \includegraphics[]{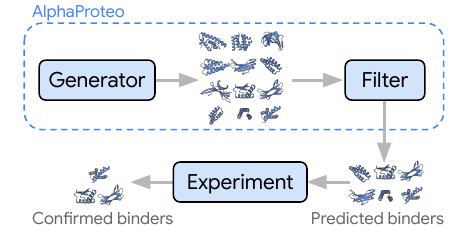}
        \phantomsubcaption\label{fig:overview:pipeline_schematic}
    \end{subfigure}
    \hfill
    \begin{subfigure}[t]{0.49\textwidth}
        \centering
        \caption*{\textbf{B}}
        \includegraphics[]{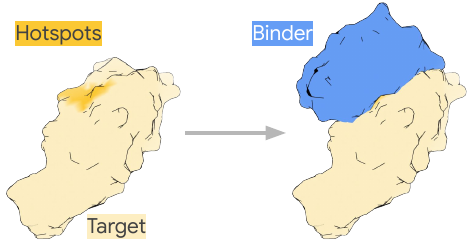}
        \phantomsubcaption\label{fig:overview:binder_design_schematic}
    \end{subfigure}
    \vfill
    \begin{subfigure}[t]{0.98\textwidth}
        \centering
        \caption*{\textbf{C}} 
        \includegraphics[]{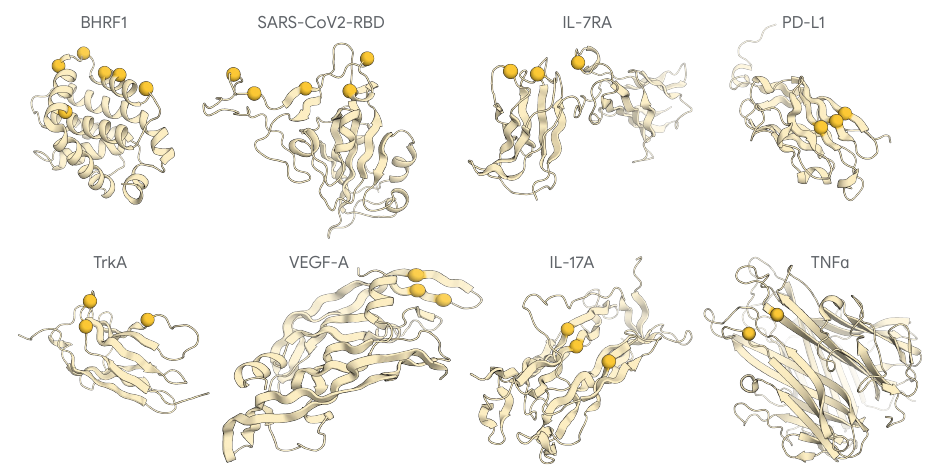}
        \phantomcaption\label{fig:overview:targets}
    \end{subfigure}
    \vfill
    \begin{subfigure}[t]{0.49\textwidth}
        \centering
        \caption*{\textbf{D}}
        \includegraphics[]{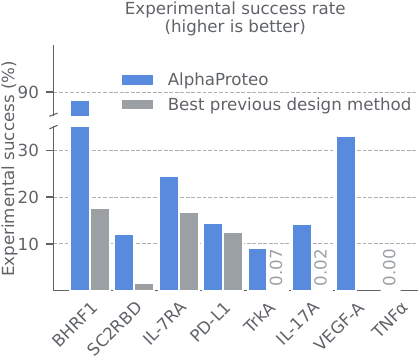}
        \phantomcaption\label{fig:overview:success_rate}
    \end{subfigure}
    \hfill
    \begin{subfigure}[t]{0.49\textwidth}
        \centering
        \caption*{\textbf{E}}
        \includegraphics[]{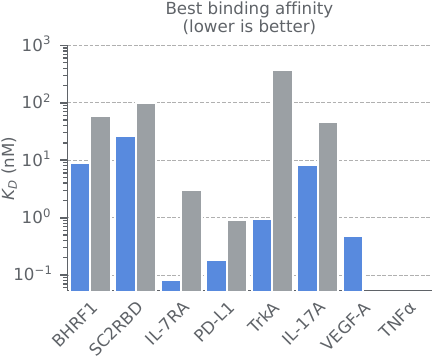}
        \phantomcaption\label{fig:overview:kds}
    \end{subfigure}
    \titledcaption{Overview and experimental performance of AlphaProteo.}{
        \textbf{(A)} Schematic of design system.
        The generative model outputs designed structures and sequences of binder candidates and the filter is a model or procedure that predicts whether a design will bind.
        \textbf{(B)} Schematic of target-structure-conditioned binder design as performed by the generative model.
        \textbf{(C)} Crystal structures (light yellow) and hotspot residues (dark yellow spheres) of seven target proteins for binder design experiments in this work.
        VEGF-A and IL-17A are both disulfide-linked homodimers.
        See \autoref{tab:binder_specs} for PDB IDs and hotspot residue numbers.
        \textbf{(D)} Percent of all tested designs with measured binding, from AlphaProteo (blue) or the best previous binder design method (gray).
        \textbf{(E)} Binding affinities of the best per-target {\KD} values from AlphaProteo (blue) or the best previous method. These represent the affinities of non-optimized computational designs -- see \autoref{tab:experimental_results} for {\KD} values of the best optimized computational designs from the literature.
        The exact values plotted in \textbf{(D)} and \textbf{(E)} are also shown in \autoref{tab:experimental_results} with data sources (see also \autoref{sec:comparison_to_other_methods}).
    }
    \label{fig:overview}
\end{figure}

To validate AlphaProteo experimentally, we designed binders against eight target proteins with diverse structural properties, of which two are viral proteins involved in infection and six are therapeutically important human proteins (\autoref{fig:overview:targets}, \autoref{tab:binder_specs}):
\newline 
\begin{enumerate}
    \item \textbf{BHRF1}, an oncogenic protein from Epstein-Barr virus; inhibition via binding can kill cancer cells and slow tumor growth~\citep{Procko2014-az}.
    It has a hydrophobic groove that perfectly accommodates a helix on its binding partner, facilitating binding.
    \item \textbf{SARS-CoV-2 spike protein receptor-binding domain (SC2RBD)}, a protein domain required for COVID-19 infection.
    We targeted its interface to the human ACE2 receptor as disrupting this interaction is known to block SARS-CoV-2 from infecting human cells~\citep{Walls2020-gx}.
    Previous design efforts have succeeded against this polar and convex site but required experimental optimization to achieve high affinity~\citep{Cao2020-cy, Gainza2023-nk}.
    \item \textbf{Interleukin-7 Receptor-$\alpha$ (IL-7RA)}, a cell-surface receptor involved in lymphocyte development and a therapeutic target for acute lymphoblastic leukemia and HIV.
    We targeted the binding site of the native interleukin-7 ligand, which is moderately hydrophobic and subject to high success rates in previous design efforts~\citep{Cao2022-vw, Watson2023-fy}.
    \item \textbf{Programmed Death-Ligand 1 (PD-L1)}, a cell-surface receptor that controls immune cell proliferation and is an important therapeutic target for cancer.
    The target site is flat and difficult to bind by small molecules and smaller proteins~\citep{Gainza2023-nk, Yang2024-dq}.
    \item \textbf{Tropomyosin Receptor Kinase A (TrkA)}, a nerve growth factor receptor involved in autoimmune disease and an analgesic target for treating chronic pain. We targeted a hydrophobic pocket addressed by previous design efforts.
    Previous binding affinities were poor without experimental optimization~\citep{Cao2022-vw}.
    \item \textbf{Interleukin-17A (IL-17A)}, a secreted protein that triggers inflammation and a therapeutic target in autoimmune disease.
    We targeted the interface of IL-17A with its native receptor, which comprises two chains of a homodimer and has a large polar pocket.
    Existing designed binders to IL-17A have poor unoptimized affinities and required screening large libraries to obtain~\citep{Berger2024-wf}.
    \item \textbf{Vascular Endothelial Growth Factor A (VEGF-A)}, a secreted growth factor controlling angiogenesis and a therapeutic target for cancer and diabetic retinopathy.
    We targeted a small hydrophobic patch bound by the native VEGF receptor~\citep{Muller1998-bv}.
    No designed binders to this target have been published despite its biomedical importance.
    \item \textbf{Tumor Necrosis Factor Alpha (TNF$\alpha$)}, a pro-inflammatory cytokine produced during inflammation and a therapeutic target for inflammatory disease \citep{Hu2013-zm,Mukai2010-ka,McMillan2021-db}. We targeted a polar region between two subunits of the TNF$\alpha$ homotrimer where it interacts with the native TNF receptor. No computationally designed binders against this target have been reported.
\end{enumerate}

We chose the above targets for their biological importance, to span a range of design problem difficulty, and to allow comparison to existing design methods.
To compare to RFdiffusion~\citep{Watson2023-fy}, we selected the target where it had the highest experimental success rate (IL-7RA) and the two targets where it had the lowest (PD-L1, TrkA), omitting the other 2 tested targets to conserve our experimental bandwidth.
We chose BHRF1 and SC2RBD as an additional easy and difficult target, respectively, which have precedent in the computational design literature.
IL-17A and VEGF-A were selected as difficult targets that had no confirmed computationally designed binders at the time of the work.
After experimental testing on the above 7 targets was completed, TNF$\alpha$ was chosen as an 8th very difficult target based on \emph{in silico} analysis (\autoref{sec:multiple_binding_hits_per_target},  \autoref{sec:in_silico_target_selection}).
No additional targets beyond these 8 were experimentally evaluated during the course of this work.

\begin{table}[t]
    \small
    \centering
    \renewcommand\arraystretch{2}
    \begin{threeparttable}
        \begin{tabular}{@{}ll*{8}{>{\centering\arraybackslash}p{0.08\linewidth}}@{}}
            \toprule
             & BHRF1 & SC2RBD & IL-7RA & PD-L1 & TrkA & IL-17A & VEGF-A & TNF$\alpha$\\
             \midrule
             & \multicolumn{8}{c}{\makecell[c]{\textbf{Experimental success rate} (\%) \\\footnotesize{(higher is better)}}}\\
             AlphaProteo & \mcb{88}{94} & \mcb{12}{172} & \mcb{25}{94} & \mc{15}{159} & \mcb{9}{131} & \mcb{14}{63} & \mcb{33}{94} & \mc{0}{54}\\
             RFdiffusion & \nan & \nan & \mc{17}{95} & \mc{13}{95} & \mc{0.0}{95} & \nan & \nan  & \nan\\
             \makecell[lt]{Other design\\ methods} & \mc{18 \tnote{a}}{17}& \mc{1.6 \tnote{b}}{63} & \mc{0.15 \tnote{c}}{14,912} & \mc{13 \tnote{b}}{16}& \mc{0.07 \tnote{c}}{14,982} & \mc{0.02 \tnote{d}}{15,000} & \nan  & \nan\\
             \midrule
             & \multicolumn{8}{c}{\makecell[c]{\textbf{Binding $\mathrm{K_D} \mathrm{(nM)}$}\\\footnotesize{(lower is better)}}}\\
             AlphaProteo & \mcb{8.5}{94} & \mc{26}{172} & \mcb{0.082}{94} & \mcb{0.18}{159} & \mcb{0.96}{131} & \mc{8.4}{63} & \mcb{0.48}{94} & \nan\\
             RFdiffusion & \nan & \nan  & \mc{14\tnote{*}}{95} & \mc{1.6\tnote{*}}{95} & \mc{370\tnote{*}}{95} & \nan & \nan & \nan \\
             \makecell[lt]{Other design\\ methods} & \mc{58 \tnote{a}}{17} & \mc{100 \tnote{e}}{100,000} & \mc{3 \tnote{c}}{14,912} & \mc{0.9 \tnote{b}}{16} & \mc{3000 \tnote{c}}{14,982} & \mc{47 \tnote{d}}{15,000} & \nan & \nan\\
             \makecell[lt]{Other design\\ methods, optimized} & \makecell{16\tnote{*, a}} & \makecell{{\bfseries 16}\tnote{*, e}} & \makecell{0.31 \tnote{c}} & \makecell{0.65 \tnote{f}} & \makecell{1.4 \tnote{c}} & \makecell{{\bfseries 0.01} \tnote{d}} & \nan  & \nan\\
             \bottomrule
        \end{tabular}
        {
            \footnotesize
            $^\textrm{a}$~\citet{Procko2014-az}
            $^\textrm{b}$~\citet{Gainza2023-nk}
            $^\textrm{c}$~\citet{Cao2022-vw}
            $^\textrm{d}$~\citet{Berger2024-wf}
            $^\textrm{e}$~\citet{Cao2020-cy}
            $^\textrm{f}$~\citet{Yang2024-dq}
        }
        \setlength{\abovecaptionskip}{8pt}
        \titledcaption{Experimental success rates and affinities of AlphaProteo and other methods.}{
            Percentage of designs with measured binding and best per-target binder affinity for AlphaProteo, RFdiffusion (as measured by us using yeast display, see \autoref{sec:comparison_to_other_methods}), and other computational design methods.
            Number of designs tested are in parentheses.
            "Other design methods, optimized" lists the best affinity after experimental optimization of any computationally designed binder.
            Binders derived from selection-based methods, such as antibodies and nanobodies, are not considered here.
            {\KD} values from the literature come from biolayer interferometry (BLI) or surface plasmon resonance (SPR) assays, except where noted by asterisks (*), where we measured the {\KD} ourselves using HTRF (\autoref{sec:comparison_to_other_methods}).
            Some targets used for method development (\autoref{sec:in_silico_benchmark}) have more detailed results in \autoref{tab:internal_model_comparison}.
        }
        \label{tab:experimental_results}
    \end{threeparttable}
\end{table}

\subsubsection{Multiple binding hits within one 96-well plate of designs per target}
\label{sec:multiple_binding_hits_per_target}
For each target, we generated a large set of \emph{in silico} designs 50-140 amino acids long (\autoref{tab:binder_specs}) and used an automated filtering procedure to choose between 47 and 172 binder candidates to test for binding by yeast surface display.
We tested designs for the initial set of seven targets and observed experimental success rates, or the fraction of designs with measurable binding (\autoref{sec:ysd}), ranging from 9\%, on TrkA, to 88\%, on BHRF1 (\autoref{tab:experimental_results}).
Per-target success rates were >5\% for 7 targets, >10\% for 6 targets and >20\% for 5 targets (\autoref{fig:overview:success_rate}, \autoref{tab:experimental_results}).

Our success rates are higher than the best alternative current method on 7 targets (\autoref{fig:biochemical_characterization:htrf}, \autoref{tab:experimental_results}). On VEGF-A, AlphaProteo is the first computational design method, to our knowledge, to obtain successful binders, although antibodies have been developed using traditional methods~\citep{Lien2008-iq}.
On BHRF1, SC2RBD, and IL-17A, AlphaProteo has, respectively, 5-, 8-, and 700-fold higher success rates than the next-best method (\autoref{fig:overview:success_rate}, \autoref{tab:experimental_results}).

To compare AlphaProteo quantitatively to RFdiffusion~\citep{Watson2023-fy}, the current state-of-the-art (SoTA) binder design method, we tested published RFdiffusion binder designs for IL-7RA, PD-L1, and TrkA alongside AlphaProteo designs in the same yeast display assay (\autoref{sec:comparison_to_other_methods}).
In this direct comparison, AlphaProteo had higher overall experimental success rates on all three targets (\autoref{fig:overview:success_rate}, \autoref{tab:experimental_results}).
These results indicate that AlphaProteo is strongly competitive to SoTA in terms of success rates.

We note that SC2RBD, PD-L1, and TrkA were used to develop AlphaProteo (\autoref{sec:in_silico_benchmark}), so these success rates may overestimate performance on novel targets.
However, for BHRF1, IL-7RA, VEGF-A, and IL-17A, we only performed a single round of medium-throughput testing, showing that high success rates can be obtained prospectively for even quite challenging targets.

After obtaining results on these seven targets, we investigated the potential target range of AlphaProteo by computing its \emph{in silico} success rate for 3 epitopes on each of 200 randomly selected target proteins from the PDB (\autoref{sec:in_silico_target_selection}).
The above 7 targets spanned a similar range of \textit{in silico} success rate as this wider list of targets, confirming that they are representative of the difficulty of most potential targets.
The screening also identified several particularly challenging targets, including TNF$\alpha$, with \textit{in silico} success rates very close to 0.
Given TNF$\alpha$'s unusual \emph{in silico} difficulty and high biomedical importance, we designed and experimentally tested binders to this target, but failed to obtain hits.
This is consistent with the low \emph{in silico} performance on this target, and is likely due to a flat, highly polar binding site at an interface between 2 subunits in a homotrimer.
Encouragingly, however, 80\% of the sampled PDB targets have higher \textit{in silico} success rates than the most difficult target where we successfully obtained binders, IL-17A (\autoref{fig:in_silico_target_histogram}).
This suggests that AlphaProteo can generalize to a wide range of biologically important binder design problems.

\subsubsection{State-of-the-art binding affinities on 7 targets}
\begin{figure}[H]
    \centering
    \begin{subfigure}[t]{0.22\textwidth}
        \centering
        \caption*{\textbf{A}}
        \includegraphics[]{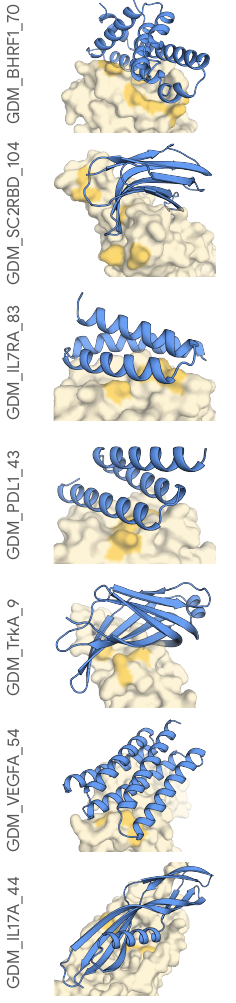}
        \phantomsubcaption\label{fig:biochemical_characterization:structures}
    \end{subfigure}
    \hfill
    \begin{subfigure}[t]{0.22\textwidth}
        \centering
        \caption*{\textbf{B}}
        \includegraphics[height=18cm]{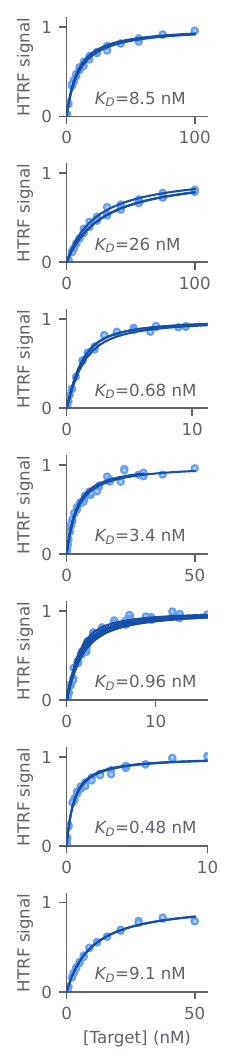}
        \phantomsubcaption\label{fig:biochemical_characterization:htrf}
    \end{subfigure}
    \hfill
    \begin{subfigure}[t]{0.3\textwidth}
        \centering
        \caption*{\textbf{C}}
        \includegraphics[height=18cm]{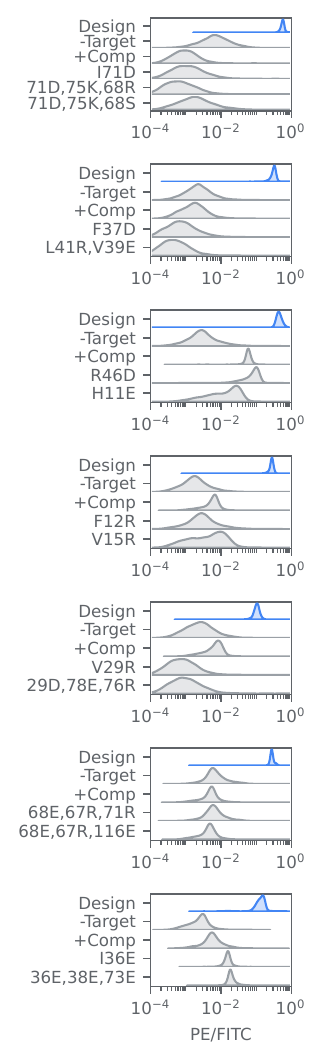}
        \phantomsubcaption\label{fig:biochemical_characterization:mutants_and_comp}
    \end{subfigure}
    \hfill
    \begin{subfigure}[t]{0.24\textwidth}
        \centering
        \caption*{\textbf{D}}
        \includegraphics[height=18cm]{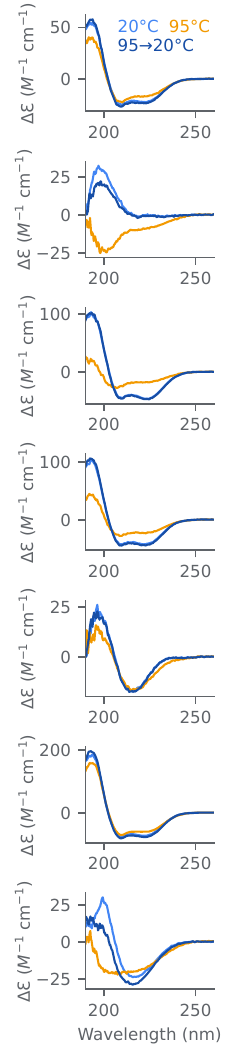}
        \phantomsubcaption\label{fig:biochemical_characterization:cd}
    \end{subfigure}
    \titledcaption{Biochemical characterization of representative binders for each target.}{
        \textbf{(A)} Design models, \textbf{(B)} HTRF equilibrium saturation binding and {\KD} values fitted from 1:1 binding models, \textbf{(C)} Yeast display on interface mutants and competitive inhibition, and \textbf{(D)} Circular dichroism spectra before (20 ºC) and after thermal melting (95 ºC and 95 $\rightarrow$ 20 ºC).
        Note that the designs here were chosen to showcase all 4 measurement types and therefore may not be the highest-affinity binder for each target.
        A list of the best binders per target and their {\KD} values can be found in \autoref{tab:sequences} (also see \autoref{fig:htrf_all}).
        HTRF y-axis is normalized to the fitted maximal signal (additional HTRF data in \autoref{fig:htrf_all}).
    }
    \label{fig:biochemical_characterization}
\end{figure}

High experimental success rates can reduce the labor and cost of obtaining binders, but once hits have been found, a far more important metric is binding affinity (\KD) to the target.
Most therapeutic antibodies have low-picomolar {\KD} values \citep{Hoogenboom2005-kd, Strohl2024-mn}, which is achieved by many rounds of experimental affinity maturation.
For binders used as research tools, low-nanomolar {\KD} values or better are also typical \citep{Landry2015-oi}.
To measure how strongly our designed binders bound their target, we recombinantly expressed and purified yeast screening hits in \emph{E. coli} to measure their {\KD} values \emph{in vitro}.
Overall, 93\% of designs chosen for follow up successfully expressed in \emph{E. coli} (\autoref{tab:successful_yeast_hits}), and the majority were monodisperse by size-exclusion chromatography (\autoref{fig:expression_yield_and_sec}).
A subset of designs assayed by circular dichroism (CD) spectroscopy all exhibited the expected secondary structures (\autoref{fig:biochemical_characterization:cd}, \autoref{fig:cd_spectra}).
Furthermore, the designs exhibited partial or no unfolding up to 95°C in CD thermal melts, indicating that they are extremely thermally stable with {\Tm} values > 95 °C (\autoref{fig:cd_spectra}).
For the recombinantly produced designs, we measured {\KD} values using a homogeneous time-resolved fluorescence (HTRF) equilibrium saturation binding assay (\autoref{sec:measuring_binding}).

AlphaProteo's best per-target {\KD} values were <1 nM for 4 targets, <10 nM for 6 targets, and <30 nM for 7 targets (\autoref{fig:overview:kds}, \autoref{tab:experimental_results}, \autoref{fig:yds_binding_signal}).
The best {\KD} overall was 82 pM, for the design IL7RA\_70 (\autoref{tab:sequences}, \autoref{fig:htrf_all}).
We identified 9 total binders with sub-nanomolar {\KD} values: 4 for IL-7RA, 2 for PD-L1, 1 for TrkA, and 2 for VEGF-A (\autoref{tab:sequences}).
Compared to the best unoptimized binders from other design methods, AlphaProteo {\KD} values were better on all targets, by margins of 7-, 4-, 37-, 5-, 380-, and 5-fold, for BHRF1, SC2RBD, IL-7RA, PD-L1, TrkA, and IL-17A, respectively (\autoref{fig:overview:kds}, \autoref{tab:experimental_results}).
Even compared to previous designed binders that have been optimized experimentally through multiple rounds of mutation and selection, the best AlphaProteo {\KD} values were still better on BHRF1, IL-7RA, PD-L1, and TrkA (\autoref{tab:experimental_results}, "Other design methods, optimized").
Taken together, the success rates and affinities achieved by AlphaProteo suggest that it can generate binders for many research applications after screening one round of 10-100 designs and no further experimentation.

\subsubsection{Designs bind the target epitope as intended}
\label{sec:intended_binding_site}
To test whether the designs bind the intended epitope on the target, we measured binding in the presence of a known competitive binder with the same target site (\autoref{fig:biochemical_characterization:mutants_and_comp}, \autoref{sec:methods}).
As expected, this reduced binding signal in all cases, with the reduction being smaller where our binders had a much higher affinity than the competitor.
To test whether our designs bind their targets via the intended interactions, we measured binding of our top binders after mutating 1-3 residues at the target-binding interface in their design models (\autoref{fig:biochemical_characterization:mutants_and_comp}, \autoref{fig:mutations_structures}, \autoref{fig:supp_mutants}).
Almost all mutants had lower binding signal than their parent, suggesting successful disruption of the binding interface by the mutations.
A small number of mutants had higher binding signal than the parent. This is not surprising given that we chose the mutations by visual intuition, which likely did not fully account for structural subtleties that could lead to improved binding (\autoref{sec:mutants_and_competition}).
Overall, these results indicate that both the binder and target interact with each other via the interfaces that were intended by design.

\subsubsection{Designs have specific binding within our target set and are structurally diverse}
\label{sec:specificity_diversity}
\begin{figure}[H]
    \centering
    \begin{subfigure}[t]{\textwidth}
        \centering
        \caption*{\textbf{A}}
        \includegraphics[trim=0 0 0 0.5cm]{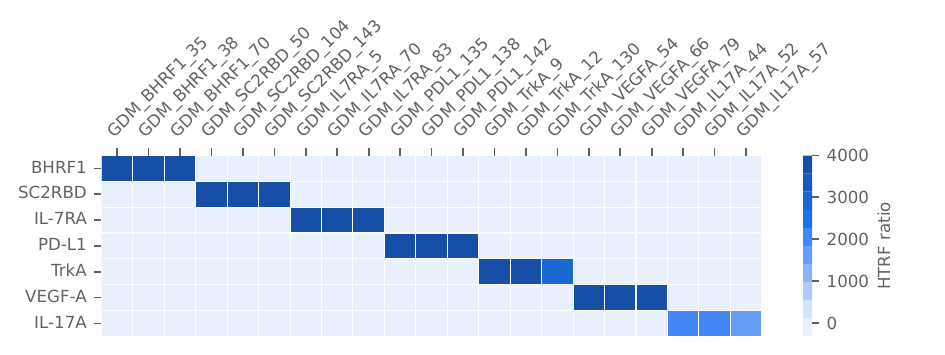}
        \phantomsubcaption\label{fig:specificity_structural_diversity:specificity}
    \end{subfigure}
    \vfill
    \begin{subfigure}[t]{\textwidth}
        \centering
        \caption*{\textbf{B}}
        \includegraphics[scale=1.05]{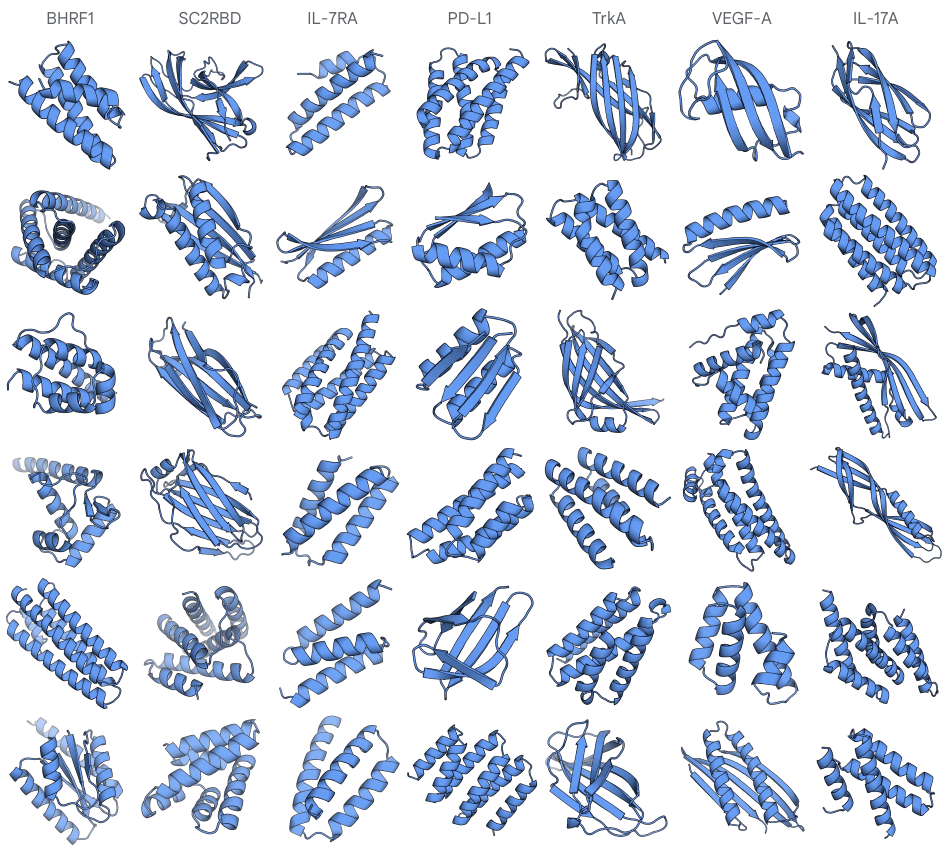}
        \phantomsubcaption\label{fig:specificity_structural_diversity:diversity}
    \end{subfigure}
    \titledcaption{Specificity and diversity of designed binders.}{
        \textbf{(A)} Specificity: HTRF binding signal of a subset of top binders (1 nM) measured against each target (100 nM).
        All binders show on-target binding signal, and none of the binders show any non-specific binding signal against any of the off-targets tested.
        \textbf{(B)} Diversity: Examples of experimentally confirmed AlphaProteo binders from different structural clusters at a TM-score cutoff of 0.6.
    }
    \label{fig:specificity_structural_diversity}
\end{figure}

To test the specificity of a subset of our top binders, we measured their binding against all 7 targets.
All binders tested exhibit observable binding only to the intended target (\autoref{fig:specificity_structural_diversity:specificity}), although it is important to note that for many downstream applications a more thorough test of specificity, such as against all proteomic targets, would need to be carried out.

We analyzed the structural diversity of our successful designs to gain insight into how many independent solutions our method is able to generate for each design problem.
Diversity is also practically important as it maximizes the chance that one of the designs will satisfy downstream requirements that are not known in advance.
We looked at the distribution of pairwise TM-scores (\autoref{fig:diversity:pairwise_tm}) and secondary structure content (\autoref{fig:diversity:secondary_structure}) across binding hits for each target.
Compared to the active designs from RFdiffusion, AlphaProteo designs were consistently lower in structural similarity to each other and had a higher frequency of all-beta structures. These observations are consistent with visual inspection of our experimentally confirmed binder designs, which reveals a variety of all-alpha, mixed alpha/beta, and all-beta folds  (\autoref{fig:specificity_structural_diversity:diversity}).

\subsection{Functional and structural validation of binders}
\label{sec:functional_structural_validation}
\subsubsection{Binders neutralize SARS-CoV-2 variants in live virus neutralization assays}
\label{sec:sarscov2_neutralization}
To determine if our binders exhibit the intended biological activity, we tested their ability to bind and neutralize live SARS-CoV-2.
We tested four of our binders (GDM\_SC2BRD\_11, GDM\_SC2BRD\_27, GDM\_SC2RBD\_104 and GDM\_SC2BRD\_50) for the ability to neutralize four variants of SARS-CoV-2 that circulated globally from 2020 and 2024 and prevent them from infecting Vero cells \citep{Shawe-Taylor2024-ko}.
All four binders successfully neutralized an ancestral strain (hCoV19/England/02/2020) with 50\% inhibitory concentrations (\EC) of 89-300 nM (\autoref{fig:functional_validation:sarscov_neutralization}, \autoref{fig:sarscov2_neutralization}).
This variant has an identical spike protein to the virus first identified in 2019 and is the source of the target structure used for design.
These {\EC} values are 2- to 10-fold higher than our measured \textit{in vitro} binding affinities (\autoref{tab:sequences}), consistent to what has been observed in the same assay for clinical monoclonal antibodies such as sotrovimab (\KD=0.21~nM, {\EC=0.67~nM} against a single SARS-CoV-2 isolate) \citep{Shawe-Taylor2024-ko}.
Interestingly, two of the binders (GDM\_SC2RBD\_11 and GDM\_SC2RBD\_129) were able to neutralize three of the tested variants.
The binder which showed the highest potency and lowest {\EC} (GDM\_SC2RBD\_50) only inhibited the ancestral variant.
All four variants were neutralized by at least one designed binder. 

\subsubsection{Binders inhibit VEGF receptor downstream signaling in cells}
\label{sec:vegfr_inhibition}
\begin{figure}[H]
    \centering
    \vspace{-5mm}
    \begin{subfigure}[t]{0.49\textwidth}
        \centering
        \caption*{\textbf{A}}
        \includegraphics[]{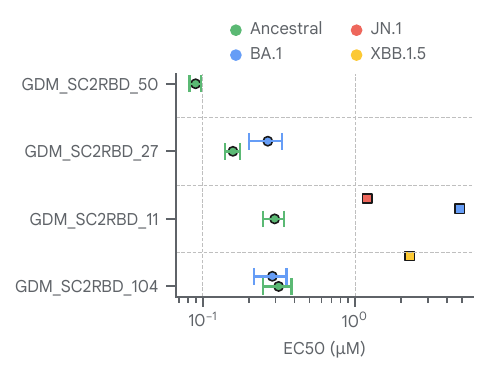}
        \phantomsubcaption\label{fig:functional_validation:sarscov_neutralization}
    \end{subfigure}
    \hfill
    \begin{subfigure}[t]{0.49\textwidth}
        \centering
        \caption*{\textbf{B}}
        \includegraphics[]{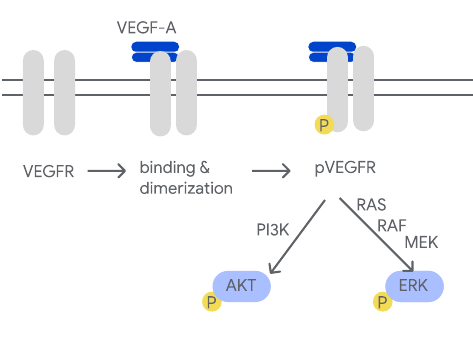}
        \phantomsubcaption\label{fig:functional_validation:binding_schematic}
    \end{subfigure}
    
    \begin{subfigure}[t]{\textwidth}
    \centering
        \vspace{-7mm}
        \caption*{\textbf{C}}
        \includegraphics[]{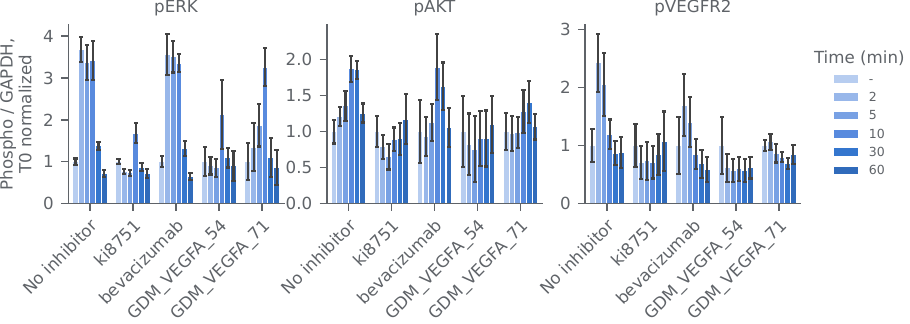}
        \phantomsubcaption\label{fig:functional_validation:vegfa_blot_quant}
    \end{subfigure}
    \begin{subfigure}[t]{\textwidth}
        \centering
        \vspace{-1mm}
        \caption*{\textbf{D}}
        \includegraphics[width=16cm]{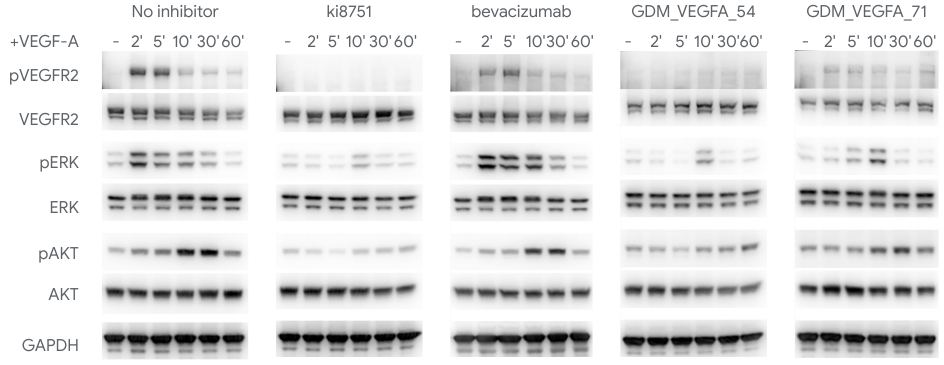}
        \phantomsubcaption\label{fig:functional_validation:vegfa_blot}
    \end{subfigure}
    \titledcaption{Inhibition of SARS-CoV-2 viral infection and VEGF signaling by designed binders.}{
        \textbf{(A)} 50\% inhibitory concentration (\EC) of 4 designed SC2RBD binders in a virus neutralization assay against 4 SARS-CoV-2 variants (\autoref{fig:sarscov2_neutralization}, \autoref{sec:methods}). Error bars show the standard error on the underlying dose-response curve. Binders with low affinity, where complete neutralisation (0\% infection) could not be observed, are displayed with square symbols. In these cases the error on the {\EC} estimate for the dose-response curves could not be meaningfully determined and error bars are omitted. 
        \textbf{(B)} Schematic representation of the VEGF-A signaling pathway.
        VEGF-A binding leads to dimerization of VEGFR, phosphorylation of VEGFR and downstream signaling cascade leading to ERK and AKT phosphorylation.
        \textbf{(C)} Ratio of phosphorylated to total ERK, AKT, and VEGFR2 western blot band intensities before (-) and 2, 5, 10, 30, and 60 minutes after treatment with small-molecule VEGFR2 inhibitor ki8751, monoclonal antibody bevacizumab, or designed VEGF-A binders. Values are normalized to pre-treatment values. Shown are the mean and S.E.M of 3 (for binders) or 6 (for controls) biological replicates.
        \textbf{(D)} Western blot of phosphorylated and total ERK, AKT, and VEGFR2 from HUVEC cells after 2 to 60 minutes of treatment with VEGF-A and binders GDM\_VEGFA\_54, GDM\_VEGFA\_71, ki8751, or bevacizumab.
        Inhibition of VEGF-A signaling is observed by a reduction in pERK, pAKT, and pVEGFR2 band intensity relative to VEGF-A-only ("no inhibitor") control.
    }
    \label{fig:functional_validation}
\end{figure}

We also tested our designed binders GDM\_VEGFA\_54 and GDM\_VEGFA\_71 for their ability to inhibit VEGF signaling.
We measured phosphorylation of VEGF receptor 2 (VEGFR2) and downstream ERK and AKT kinases in primary human umbilical vein endothelial cells (HUVECs) stimulated with human VEGF-A (\autoref{fig:functional_validation:vegfa_blot_quant}).
Incubation with GDM\_VEGFA\_54 leads to substantially reduced phosphorylation of ERK, AKT, and VEGFR2 compared to a VEGF-A-only control (\autoref{fig:functional_validation}, "no inhibitor").
This effect is similar to that of ki8751 \citep{Kubo2005-lt}, a potent small-molecule VEGFR2 kinase inhibitor.
The effect is more potent than that of the anti-VEGF-A monoclonal antibody bevacizumab, the active component of the clinically approved drug Avastin \citep{Kramer2007-ri}, which we tested at an equimolar concentration to our binders in this experiment.
This concentration of bevacizumab is 1000-fold lower than that usually tested \textit{in vitro} on HUVECs \citep{Jia2019-me}, suggesting that GDM\_VEGFA\_54 is a more potent VEGF-A inhibitor than bevacizumab in HUVECs.
The second binder tested, GDM\_VEGFA\_71, leads to a weaker, although still visible reduction in phosphorylation of ERK, AKT, and VEGFR2.
These results are consistent with our relative \textit{in vitro} binding affinities of GDM\_VEGFA\_54 and GDM\_VEGFA\_71 for VEGF-A, which are 0.48 and 4.7 nM, respectively.

\subsubsection{Experimental structures of binder-target complexes confirm binding mode and structure}
\label{sec:experimental_structures}
\begin{figure}[H]
    \centering
    \begin{subfigure}[t]{\textwidth}
        \centering
        \caption*{\textbf{A}}
        \includegraphics[]{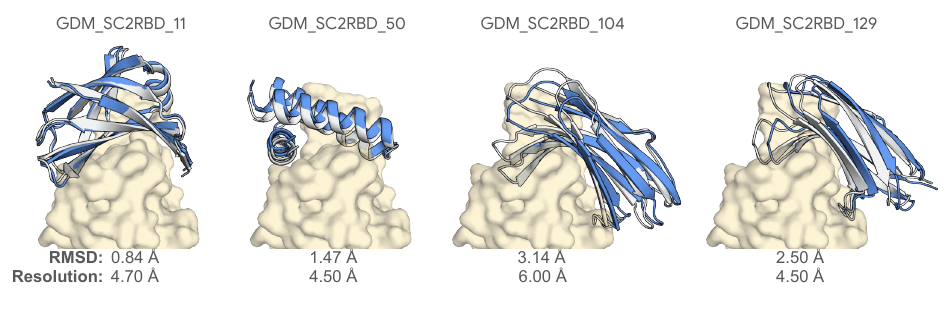}
        \phantomsubcaption\label{fig:cryoem_sarscov2_xtal_vegfa:cryoem_sarscov2}
    \end{subfigure}
    \begin{subfigure}[t]{0.56\textwidth}
        \centering
        \caption*{\textbf{B}}
        \includegraphics[]{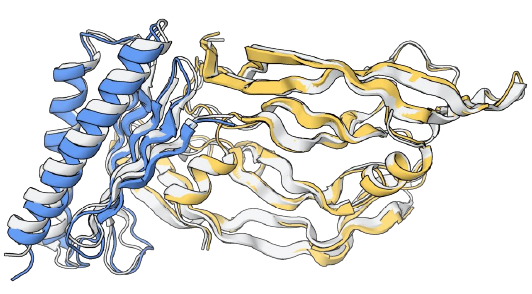}
        \phantomsubcaption\label{fig:cryoem_sarscov2_xtal_vegfa:xtal_vegfa}
    \end{subfigure}
    \hfill
    \begin{subfigure}[t]{0.43\textwidth}
        \centering
        \caption*{\textbf{C}}
        \includegraphics[]{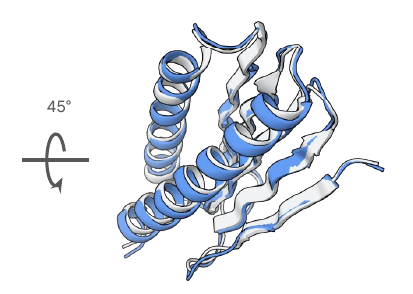}
        \phantomsubcaption\label{fig:cryoem_sarscov2_xtal_vegfa:vegfa_binder_monomer}
    \end{subfigure}
    \begin{subfigure}[t]{0.56\textwidth}
        \centering
        \caption*{\textbf{D}}
        \includegraphics[]{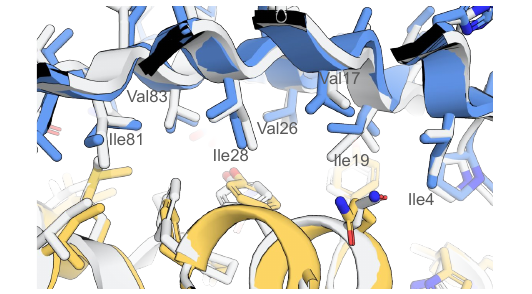}
        \phantomsubcaption\label{fig:cryoem_sarscov2_xtal_vegfa:vegfa_binder_interface1}
    \end{subfigure}
    \begin{subfigure}[t]{0.43\textwidth}
        \centering
        \caption*{\textbf{E}}
        \includegraphics[]{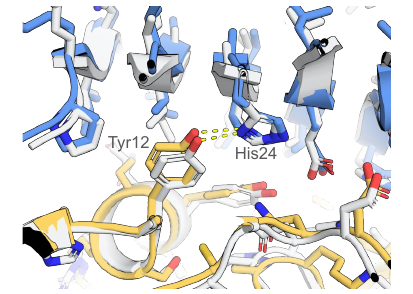}
        \phantomsubcaption\label{fig:cryoem_sarscov2_xtal_vegfa:vegfa_binder_interface2}
    \end{subfigure}
    \titledcaption{Experimental structures of binders to SARS-CoV-2 spike and VEGF-A.}{
        \textbf{(A)} Cryo-EM structures of designed binders (blue) in complex with SARS-CoV-2 spike protein (yellow), aligned to AF2-multimer prediction (gray) on spike protein.
        Values are shown for the cryo-EM structure resolution and target-aligned binder C$\alpha$ RMSDs between AF2-multimer and experimental structures.
        \textbf{(B)} Crystal structure of complex between VEGF-A homodimer (yellow) and design GDM\_VEGFA\_71 (blue), aligned to AF2-multimer prediction (gray) on VEGF-A (binder C$\alpha$ RMSD = 1.65 {\AA}).
        \textbf{(C)} Rotated view of binder monomer (binder-aligned binder C$\alpha$ RMSD = 0.78 {\AA}).
        \textbf{(D-E)} Closeup of the binder-target interface showing close agreement of sidechains between experimental structure and AF2-multimer prediction of design.
        \textbf{(D)} Packing of hydrophobic sidechains of the binder at the interface. Most have near-perfect agreement between design and structure, except Val17, Ile19, and Ile81, which have slight deviations.
        \textbf{(E)} A designed hydrogen bond between His24 of the binder and Tyr12 of VEGF-A.
    }
    \label{fig:cryoem_sarscov2_xtal_vegfa}
\end{figure}

To validate the structures and binding modes of our designs, we used cryo-electron microscopy (cryo-EM) to obtain structures of GDM\_SC2RBD\_11, GDM\_SC2RBD\_50, GDM\_SC2RBD\_104, and GDM\_SC2RBD\_129 in complex with the SARS-CoV-2 spike S1 protein at 4.5 - 6.0 {\AA} resolution (\autoref{fig:cryoem_sarscov2_xtal_vegfa:cryoem_sarscov2} and \autoref{fig:sarscov2_cyroem_data_processing}). The experimental structures closely recapitulate the designed binder-target complexes, with binder C$\alpha$ RMSDs of 0.84 - 3.14 {\AA} using the target S1 protein as an alignment reference.
We additionally obtained an X-ray crystal structure of GDM\_VEGFA\_71 in complex with VEGF-A, at 2.65 {\AA} resolution (\autoref{fig:cryoem_sarscov2_xtal_vegfa:xtal_vegfa}).
The binder folded extremely closely to its designed structure, a mixed alpha-beta fold with a 5-strand beta sheet interacting with VEGF-A, demonstrating atomic level accuracy that shows a C$\alpha$ RMSD of 0.78 {\AA} between AF3 model and experimental structure.
The designed binding orientation was also highly accurate, with a target-aligned binder C$\alpha$ RMSD of 1.65 {\AA}.
Most sidechains of the binder interacting with the target also had the correct rotamer, including a buried hydrogen bond between a histidine of the binder and a tyrosine of VEGF-A which was recapitulated almost perfectly in the experimental structure (\autoref{fig:cryoem_sarscov2_xtal_vegfa:vegfa_binder_interface2}).

\section{Conclusion}
\label{sec:conclusion}
Our results show that AlphaProteo is capable of generating low- to sub-nanomolar binders for a diverse range of targets after a single round of medium-throughput testing.
The binders are small (5-15 kDa), thermostable, and highly expressed, and therefore potentially already suitable for use in some research applications without further optimization.
However, it is important to note that we have experimentally validated relatively few targets in this work, and all our binders are designed using a target crystal structure as input.
We hope to further improve AlphaProteo's performance and expand its capabilities to address a wider range of binder design problems, including challenging targets such as TNF$\alpha$ as well as those which lack experimental structures or a single well-defined conformation.
We believe that AlphaProteo will unlock new solutions for many biological applications, such as controlling cell signaling, imaging proteins, cells, and tissues, conferring target specificity to various effector systems, and beyond.

\subsection*{Additional notes}
The contents of this report are intended for research purposes only, and not for clinical use. This report does not include machine learning methods due to biosecurity and commercial considerations. We are looking to develop a safe and responsible protein design offering for the community, informed by our work and consultations on biosecurity and safety.

\subsection*{Acknowledgements}
The authors would like to thank the following people for their input and feedback: Jonas Adler, Andy Ballard, Charlie Beattie, David Belanger, Lucy Colwell, Andrew Cowie, Sarah Elwes, Richard Evans, Conor Griffin, John Jumper, Svend Kj{\ae}r, Antonia Paterson, Matteo Perino, Francesca Pietra, Uchechi Okereke, Olaf Ronneberger, Freyr Sverrisson, Nick Swanson, Kathryn Tunyasuvunakool, Augustin Žídek. We would also like to thank Dane Wittrup (Dept.~of Chemical Engineering, Massachusetts Institute of Technology) for his generous gift of yeast vector pCTcon2 and Svend Kj{\ae}r (Structural Biology Science Technology Platform, The Francis Crick Institute) for his production of the SARS-CoV-2 spike protein. 

\subsection*{Contributions} 
Machine learning model development, generation of design candidates, experimental success rate, experimental binding affinity measurements, and VEGF-A binder crystal structure determination were performed by Google DeepMind.

Cell-based assays and cryo-EM structure determination were performed by research groups at The Francis Crick Institute, London, UK.

\clearpage
\begingroup
\sloppy 
\printbibliography[notkeyword={alphaproteo-wp-refs/methods}, heading=bibintoc]
\endgroup
\end{refsection} 

\clearpage
\appendix
\begin{refsection}  
\DeclareFieldFormat{labelnumber}{S#1}
\setlength{\labelnumberwidth}{3em}

\section*{Supplementary information}
\addcontentsline{toc}{section}{Supplementary information}

\renewcommand{\thesection}{S\arabic{section}}
\renewcommand{\thefigure}{S\arabic{figure}}
\renewcommand{\thetable}{S\arabic{table}}
\renewcommand{\theequation}{S\arabic{equation}}

\setcounter{section}{0} 
\setcounter{figure}{0}  
\setcounter{table}{0}   
\setcounter{equation}{0} 

\section{Experimental methods}
\label{sec:methods}

\subsection{Target protein expression and purification}
\label{sec:protein_expression_and_purification}
Purified protein stocks for IL-7RA(21-239), TrkA(34-423), PD-L1(19-239), VEGF-A(27-191), and IL-17A(24-155) were purchased from BioTechne, with catalog numbers AVI10317, AVI11378, AVI156, AVI293, and BT7955, respectively.
IL-7RA, PD-L1, and TrkA have C-terminal Fc and biotinylated Avi tags, while VEGF-A has a biotinylated C-terminal Avi tag and IL-17A is biotinylated via sugars.
VEGF-A and IL-17A are disulfide-linked homo-dimers. 
For X-ray crystallography, VEGF165 (Uniprot P15692-4) was purchased from Qkine, with catalog number Qk048.

For BHRF1, a recombinant protein construct (Uniprot P03182, residues 2-160) was produced with an N-terminal Twin-Strep tag and a 3C protease cleavage site.
Transformed BL21 (DE3) (Thermo Scientific) cultures were grown in Terrific Broth (TB) medium (Melford) supplemented with carbenicillin (50 µg/mL) at 37 ºC with shaking.
At OD$_{600}$ = $\sim$0.6, protein expression was induced with 0.1 mM IPTG, the temperature reduced to 21 ºC and cultures were grown overnight.

Cells were harvested and resuspended in 20 mM Tris pH 8.0, 300 mM NaCl supplemented with 0.5 mg/mL lysozyme, 100 U DNase I, 1 mM MgCl$_2$ and a cOmplete EDTA-free protease inhibitor tablet (Roche) at a 1:5 cell weight to buffer ratio. Cell lysis was achieved by sonicating the cell suspension at 40\% amplitude (15 seconds on / 45 seconds off) for 24 cycles on ice.
Lysate was centrifuged at 48,000 x \textit{g} for 45 min at 4 ºC and the supernatant was recovered and filtered through a 0.45 µm filter (Sartorius).
The sample was applied to a 5 mL StrepTrap XT column (Cytiva) pre-equilibrated with Strep binding buffer (100 mM Tris pH 8.0, 150 mM NaCl, 1 mM EDTA pH 8.0, 0.5 mM TCEP) using an AKTA Pure 25 M.
Following sample application, the column resin was washed with 10 column volumes (CV) of the same buffer before the protein was eluted with 10 CV of 1x BXT elution buffer (IBA Lifesciences) supplemented with 0.5 mM TCEP.
1 CV fractions were collected and assessed via SDS-PAGE to confirm presence of the protein of interest.
BHRF1 was pooled and concentrated using a 10 kDa MWCO concentrator (Vivaspin). The sample was further purified by size exclusion chromatography (SEC) using a Superdex 75 increase 10/300 GL column pre-equilibrated with 20 mM sodium phosphate pH 7.5, 0.5 mM TCEP.
Fractions were confirmed by SDS-PAGE and the concentration was measured by absorbance at 280 nm using a NanoDrop One (Thermo Scientific) and the BHRF1 construct’s theoretical extinction coefficient \citep{Stoscheck1990-to}.
Purified protein was aliquoted and stored at -80 °C.

For SC2RBD, a recombinant protein construct (NCBI reference NC\_045512, residues 319-541) of SARS-CoV-2 Spike S1 glycoprotein corresponding to the receptor binding domain was produced with a C-terminal Twin-Strep tag. The signal peptide from immunoglobulin kappa gene product (METDTLLLWVLLLWVPGSTGD) was used to direct secretion of the construct.
The corresponding codon-optimized DNA fragment was cloned into mammalian expression vector pQ-3C-2xStrep for expression in Expi293F cells.
Expi293F cells grown at 37 °C in 5\% CO$_2$ in shake flasks containing FreeStyle 293 medium were transfected with endotoxin free plasmid preparation using ExpiFectamine reagent (Thermo Fisher Scientific). Conditioned medium was harvested 4 and 8 days post-transfection.
Recombinant protein was captured on Streptactin XT (IBA LifeSciences) affinity resin. Following extensive washes in TBSE buffer (20 mM Tris-HCl pH 8.0, 150 mM NaCl, 1 mM EDTA), the protein was eluted in 1x BXT buffer (IBA LifeSciences) and further purified by SEC using a Superdex 200 16/600 column (GE Healthcare) in TBSE buffer.
The purified protein was concentrated using a 10 kDa MWCO concentrator (Sartorius), aliquoted, snap-frozen in liquid nitrogen and stored at -80 °C.

\subsection{Yeast surface display and flow cytometry}
\label{sec:ysd}

\subsubsection{Primary binding screen}
Binder design sequences were codon-optimized by DNAworks \citep{Hoover2002-is} and most were synthesized by Twist as gene fragments flanked by \emph{BsaI} restriction sites as well as homology regions to a modified pETcon vector (pCTcon2, a generous gift from K.
Dane Wittrup, MIT).
\emph{Saccharomyces cerevisiae} strain EBY100 cells (50 $\mu$L) were transformed using a modified lithium acetate method without using the single-strand carrier DNA \citep{Gietz2007-bo} with 50 ng of linearised plasmid and a minimum of 10 ng of gene fragment insert in a 96-well plate.
Cells were grown at 30 ºC shaking at 1,000 rpm in complete synthetic medium -Trp -Ura + 2\% glucose for 48-72 hours.
For protein expression, a volume of yeast cells were centrifuged at 1,800 x g for 5 minutes at 20 °C and resuspended in 1 mL complete synthetic medium + 0.1\% glucose + 2\% galactose (SGCAA) to OD$_{600}$ = 1.0.
Cells were incubated at 30 °C overnight and a volume of cells at OD$_{600}$ = 0.4 were washed twice with 200 µL of 1x PBS + 0.1\% BSA (PBSF), centrifuged at 1,800 x \textit{g} for 3 minutes at 20 °C and the supernatant was removed.

To screen for binding, yeast cells were then incubated with biotinylated target proteins (diluted in PBSF) for 1 hour, washed twice with PBSF and incubated with 25 µg/mL fluorescein isothiocyanate (FITC)-conjugated anti-Myc antibody (FITC-Ab) (Abcam) and 30 µg/mL streptavidin-phycoerythrin (SAPE, Thermo Fisher Scientific) for 30 minutes. For VEGF-A and IL-17A, an avidity method with increased sensitivity was used; target proteins were pre-incubated with 25 µg/mL FITC-Ab and 30 µg/mL SAPE for 30 minutes before incubating with cells.
Following binding, cells were washed once with PBSF and resuspended in 200 µL of PBSF. Cells were analyzed on the CytoFlex LX (Beckman Coulter) or ZE5 Cell Analyzer (Bio-Rad) flow cytometers by measuring fluorescence of FITC and phycoerythrin (PE) to detect binder expression and target binding respectively.

\begin{figure}[H]
    \centering
    \begin{subfigure}[t]{0.33\textwidth}
        \centering
        \caption*{\textbf{A}}
        \includegraphics[width=5.75cm]{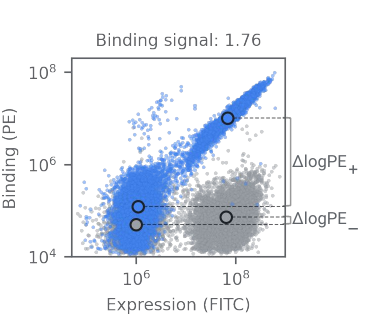}
        \phantomsubcaption\label{fig:yds_binding_signal:hit}
    \end{subfigure}
    \hfill
    \begin{subfigure}[t]{0.66\textwidth}
        \centering
        \caption*{\textbf{B}}
        \includegraphics[width=5cm]{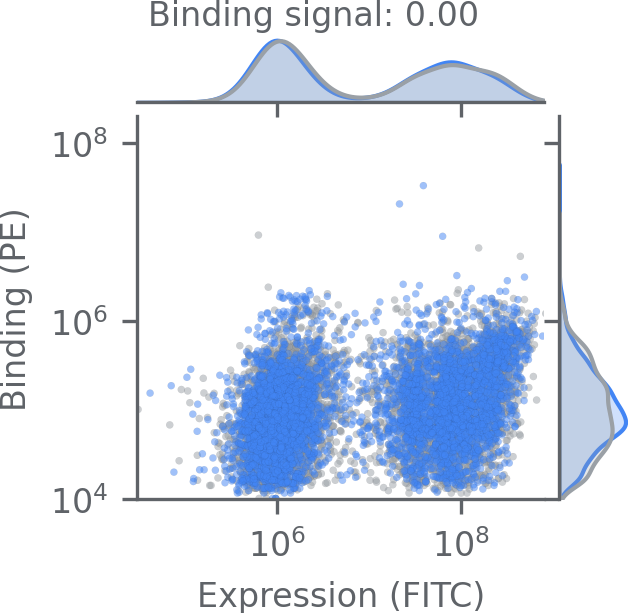}
        \includegraphics[width=5cm]{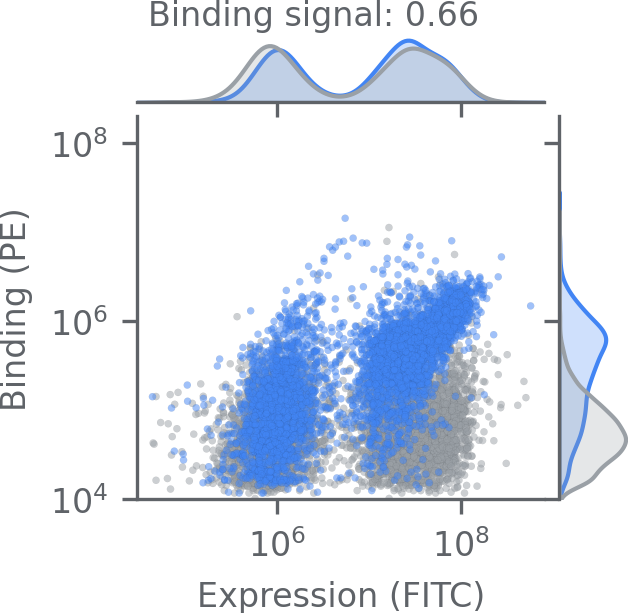}
        \phantomsubcaption\label{fig:yds_binding_signal:no_and_weak_hit}
    \end{subfigure}
    \vspace{-1.2cm}
    \vfill
    \begin{subfigure}[t]{\textwidth}
        \centering
        \caption*{\textbf{C}}
        \includegraphics[]{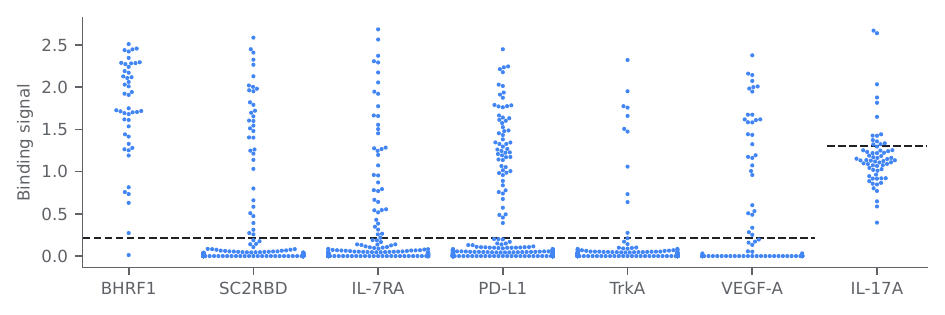}
        \phantomsubcaption\label{fig:yds_binding_signal:summary}
    \end{subfigure}
    \titledcaption{Yeast display screening of binder designs.}{
        \textbf{(A)} Binding signal computed as $(\Delta \log(\mathrm{PE}_{+}) - \Delta \log(\mathrm{PE}_{-}))$ was used to systematically determine binding success.
        This metric captures the shift in PE signal for the positive population (binding in the presence of the target) in excess of PE shift in the negative population (in the absence of the target), which factors out experiment artifacts which could lead to false positives.
        \textbf{(B)} Examples of FITC/PE scatterplots for no binding (left) and weak binding (right).
        \textbf{(C)} Binding signal distribution by target for designs tested via yeast surface display.
        Dotted lines denote the binary binding threshold: for the first 6 targets the cut-off is set to 0.2 and was empirically determined by FITC/PE plot analysis, whereas for IL-17A the cut-off was set more stringently (due to anomalous yeast display behavior), based on its positive binding control.
    }
    \label{fig:yds_binding_signal}
\end{figure}

Flow cytometry data were analyzed to compute a "binding signal", defined as:
\begin{align}
\text{signal} = (\log_{10}\PE{+}{+} - \log_{10}\PE{-}{+}) - (\log_{10}\PE{+}{-} - \log_{10}\PE{-}{-}) \nonumber
\end{align}
where $\PE{+}{+}$ is the mean PE (binding) signal of the FITC+ (binder-expressing) subpopulation in a well where target protein has been added (\autoref{fig:yds_binding_signal:hit}, \autoref{fig:yds_binding_signal:no_and_weak_hit}).
"FITC-" indicates the non-binder-expressing cell population, and "-target" indicates a control well containing the same binder but to which no target has been added. FITC+ and FITC- cells are identified by k-means clustering.
This metric captures the shift in PE signal due to binder expression and target binding in excess of the PE shift due to binder expression alone or target binding alone, thus controlling for experiment artifacts which could lead to false positives. Designs with a binding signal > 0.2 were considered successful binders, except in the case of IL-17A, where this threshold was set to 1.3 to account for background binding.
These thresholds were calibrated manually by visually inspecting scatterplots of the raw yeast data.

\subsubsection{Interface mutation, competitive inhibition, and specificity experiments}
\label{sec:mutants_and_competition}
Interface mutations were selected by manual visual inspection of predicted structures of the designed binder-target complexes.
We generated single-mutants with a hydrophobic residue (alanine, valine, leucine, and isoleucine) on the target-facing interface of the binder changed to a charged residue (aspartate, glutamate, arginine, lysine), as well as a small number of multiple-mutants with combinations of the single mutations.
Mutants were screened following the same method as the primary binding screen.

For competition assays, the competitor protein used for BHRF1, SC2RBD, IL-7RA, PD-L1, VEGF-A, and IL-17A, respectively, are BINDI \citep{Procko2014-az}, LCB1 \citep{Cao2020-cy}, RFD\_IL7RA\_55, RFD\_PDL1\_76, RFD\_TrkA\_88~\citep{Watson2023-fy}, VEGFR1 (ACROBiosystems, VE1-H52H9) and IL-17R (Biotechne, 11234-IR-100).
Yeast cells were incubated with biotinylated target proteins with or without a competitor protein for 1 hour (the competitor protein was added to the biotinylated target protein master mix just before adding to the cells).
The cells were then washed twice with PBSF and incubated with 25 µg/mL FITC-Ab (Abcam) and 30 µg/mL SAPE (Thermo Fisher Scientific) for 30 minutes.
For VEGF-A and IL-17A, an avidity method with increased sensitivity was used, similar to the primary binding screen; target proteins were pre-incubated with 25 µg/mL FITC-Ab and 30 µg/mL SAPE for 30 minutes before adding competitor protein and incubating with cells.
Following binding, cells were washed once with PBSF and resuspended in 200 µL of PBSF.

To test for specificity, 1 nM of each binder was tested for binding against 100 nM target using a homogeneous time resolved fluorescence (HTRF) assay readout and similar methods to those described in the HTRF methods section below.

\subsection{Designed binder expression and purification}
Designed binders with the highest binding signal by yeast display (\autoref{fig:yds_binding_signal}) were selected for \emph{E. coli} expression and follow up experiments.
Designs purchased as gene fragments were cloned into a modified pTriEx-4 vector containing an N-terminal 8-His tag and a 3C protease cleavage site using NEBridge Golden Gate cloning (NEB) at \emph{BsaI} sites, transformed into DH5-$\alpha$ competent cells (Thermo Scientific), miniprepped (Qiagen), and verified by Sanger sequencing (Azenta).
A small number of designs were purchased from Twist Bioscience directly as cloned plasmids in pTriEx-4 or pET-29b.
For expression, plasmids were transformed into BL21 (DE3) cells and the entire transformation mix inoculated into autoinduction medium consisting of TB medium, 0.05\% glucose, 0.2\% alpha-lactose, and 50 µg/mL carbenicillin or 50 µg/mL kanamycin.
Cultures were incubated at 37 °C with shaking (220 or 1000 rpm) for 24 hours, harvested at 2,568 x \textit{g} for 10 minutes, and pellets stored at -80 °C until purification.
Cell pellets were chemically lysed using BugBuster Master Mix (Novagen) supplemented with cOmplete EDTA-free protease inhibitor (Roche) with shaking for 20 minutes at room temperature.
Lysates were clarified by centrifugation for 1 hour at 2,568 x \textit{g}, then purified by immobilized metal affinity chromatography (IMAC) using Ni-NTA in either 0.1 mL spin columns (Cytiva) or in HisPur™ Ni-NTA 96-well Spin Plates (Thermo Scientific), followed by SEC on an AKTA Pure 25 M (Cytiva) equipped with an ALIAS autosampler (Spark Holland) using a Superdex 75 increase 10/300 GL column equilibrated in 20 mM sodium phosphate pH 7.5.
Protein samples were analyzed by SDS-PAGE, and where required, concentrated using a 3 kDa MWCO Vivaspin concentrator (Cytiva).
Protein concentrations were measured in triplicate by absorbance at 280 nm with a NanoDrop One (Thermo Scientific) using theoretical extinction coefficients \citep{Stoscheck1990-to}.
Binders that have no theoretical extinction coefficients were assay (BCA assay, Thermo Scientific) or Bradford assay (Thermo Scientific). Purified proteins were aliquoted and stored at -80 °C until further use.

Where larger quantities of designed binders were required, for example for CD and X-ray crystallography experiments, expression was scaled up to 100-1000 mL BL21 (DE3) cultures and the above protocol followed with minor modifications. Cells were lysed using sonication on ice, and lysates were clarified by centrifugation at 48,000 x \textit{g} for 45 minutes at 4 ºC before being applied to 5 mL Ni-NTA column (Cytiva), followed by SEC.

\subsection{Measurement of binding affinity / binding dissociation constants (\KD)}
\label{sec:measuring_binding}

\subsubsection{Homogeneous Time Resolved Fluorescence (HTRF)}
\label{sec:htrf}
Binding affinities ($\mathrm{K_D}$s) were measured in equilibrium saturation-binding experiments with fixed binder design concentration and target titration. The total assay volume was 16 µL and all proteins and reagents were diluted in PPI europium detection buffer (Revvity).
Target protein was premixed with HTRF acceptor reagent Streptavidin-d2 (Revvity), serially diluted, and transferred to a white ProxiPlate 384-shallow well microplate (‘assay plate’, Revvity).
Subsequently, 1 nM of each binder was added to the assay plate in duplicate (binders with {\KD} < 0.5 nM were later re-assayed with 0.1 nM binder to ensure robust data fitting).
The assay plate was centrifuged at 500 x \textit{g} for 30 seconds, sealed and incubated at room temperature for between 30 minutes and 1 hour.
HTRF donor mAb Anti-6HIS-Eu Gold (Revvity) was then added to a final concentration of 2 nM (1x), using a Mantis microfluidic liquid dispenser (Formulatrix) running software version 5.1.1 on Windows 10. The assay plate was centrifuged, sealed and incubated for a further 1 hour at room temperature.
HTRF signal was measured using a PHERAstar FSX (BMG) plate reader equipped with an HTRF 337/665/620 optic module running software version 5.70 R6 on Windows 10.
The measurement conditions were as follows; 60 µs integration delay, 400 µs integration time, 60 flashes.
The optimal focal (Z) height was determined using channel B for each experiment.
HTRF ratios were calculated by dividing the acceptor signal at 665 nm by the donor signal at 620 nm, and multiplying by a factor of 10,000.
Mean background signal for each target-acceptor concentration (0 nM binder) was subtracted, and data were analyzed using custom Python code by fitting to the general 1:1 binding equation.
\begin{align}
    R = \frac{R_{\text{max}}}{2B}\left(B + A + K_D - \sqrt{(B + A + K_D)^2 - 4AB)}\right) \nonumber
\end{align}
where R is the measured equilibrium HTRF signal, A and B are the titrated and fixed binding partner concentrations, respectively, and Rmax and $\mathrm{K_D}$ are the fitted maximal HTRF signal and binding dissociation constants, respectively. We used this equation because some of our binders had $\mathrm{K_D}$ values close to or lower than the fixed binder concentration used in the experiment, which causes the more common hyperbolic equation of 1:1 binding to overestimate the true $\mathrm{K_D}$ \citep{Jarmoskaite2020-eu}. To ensure reliable model fitting, we always used a fixed binder concentration no more than 2-fold higher (and usually much lower) than the estimated $\mathrm{K_D}$ \citep{Jarmoskaite2020-eu}.

\subsubsection{Bio-Layer Interferometry (BLI)}
\label{sec:bli}
For selected controls and designs, we measured $\mathrm{K_D}$s by kinetic BLI assays to establish confidence in the HTRF results (see "Comparison of binding affinity (\KD) to other methods" above).
Data were collected on the Octet R8 (Sartorius AG, Göttingen, Germany) using the integrated Octet Discovery software version 12.2.2.20.
Recombinant proteins were diluted from concentrated frozen stocks in 20 mM sodium phosphate pH 7.5, 0.05\% Tween-20 (BLI buffer). 
A seven-point dilution series of the analyte protein was also prepared in BLI buffer to create a titration curve. Ni-NTA biosensors (Sartorius, catalog number 18-5102) were prequilibrated in BLI buffer for at least 10 minutes prior to starting the experiment.
A fixed concentration of "ligand" (8His-tagged binder) was loaded onto sensors for 120-240 seconds, briefly washed for 10 seconds, followed by a 60 second baseline.
Association of a titration series of "analyte" (target protein) was then performed for 90-420 seconds, followed by dissociation for 600-1200 seconds.
All steps were performed at 25 °C and with shaking at 1000 rpm. Loading, association, and dissociation durations were optimized for each binder-target pair. Data were processed using Octet Analysis Studio (version 12.2.2.26).
Measurements from reference sensors not loaded with ligand, as well as a reference well with 0 nM analyte, were subtracted from the final data to account for non-specific binding of analyte to the sensors and baseline drift due to unloading of ligand from sensors, respectively.
Baseline (pre-association) signal was aligned to 0 before final analysis, where kinetic constants were obtained by nonlinear regression of 1:1 or 2:1 binding equations to the data.
Fits were performed globally, over both association and dissociation, with a shared R$_\text{max}$ for all analyte concentrations.

\subsection{Circular dichroism (CD) spectroscopy}
Data were collected on a Jasco J-815 circular dichroism spectrometer, running software Spectra Manager Version 2.15.20, equipped with a PTC-348 temperature control device.
Far-UV spectra (260-190 nm) and thermal unfolding measurements were recorded in 1 mm quartz glass cuvettes (Hellma) containing protein solutions at 10 µM in 20 mM sodium phosphate pH 7.5. Baselines containing 20 mM sodium phosphate pH 7.5 were collected prior to sample analysis.

Spectra were recorded in the far-UV (260-190 nm) at 20 °C with a scanning speed of 200 nm/min and a digital integration time (DIT) of 0.25 seconds.
25 accumulations (spectral scans) were recorded and automatically averaged by the software. Thermal unfolding data were recorded at 222 nm between 2-95 °C at a ramp rate of 2 °C/min, with measurements recorded at 0.2 °C intervals.
The DIT was set to 4 seconds. Following thermal unfolding measurements, spectra in the far-UV were collected at 95 °C to measure CD spectra changes post thermal unfolding.
Additional CD spectra were then collected following cooling of the same samples to 20 °C, to observe refolding.

Data processing: For spectra, the buffer baseline scans were subtracted from each sample dataset, the final 15 nm of measurements (between 260-245 nm) were normalized, and the CD signals in mdeg were converted to {$\mathrm{\Delta\varepsilon (M^{-1} cm^{-1})}$}.

\subsection{Western blot analysis of VEGF-A signaling in HUVECs}
Human umbilical vein endothelial cells (HUVEC, PromoCell \#C-12008) were cultured in ECG Medium 2 KIT (PromoCell, \#C-22111) according to the manufacturer's instructions.
Cells were plated at passage 6 in a six well format.
The following day cells were starved for 4 hours in Endothelial Cell Basal Medium 2 (PromoCell, \#C-22211).
Control inhibitors or binder proteins were added after 3 hours of starvation for the remaining hour at the following concentration: 1 µM ki8751 (Bio-Techne, \#2542/10), 1 µM bevacizumab (Biosynth, \#FB76708), or 1 µM binders GDM\_VEGFA\_54, GDM\_VEGFA\_71.
Following the 1 hour treatment, cells were stimulated with 30 ng/mL hVEGF-A (Peprotech, \#100-20-10UG) for 2, 5, 10, 30 or 60 minutes (0 minute refers to untreated). Subsequently, cells were washed with ice cold PBS and frozen at -80 °C.

Cells were lysed in 60 µl/well D0.4 lysis buffer (20 mM HEPES pH 7.5, 0.4 M NaCl, 10\% glycerol, 0.4\% Triton X-100, 10 mM EGTA and 5 mM EDTA, 1x HALT protease inhibitor (Thermo Fisher Scientific, \#87786), 1x HALT phosphatase inhibitor (Thermo Fisher Scientific, \#78420), 1 mM DTT, 25 mM NaF and 25 mM sodium-b-glycerophosphate. The protein concentration of each sample was measured using the Pierce™ BCA assay (Thermo Fisher Scientific, \#10678484) and the protein concentration from each replicate was adjusted to the same concentration per sample.

Western blots were performed using 4-12\% Bis-Tris SDS-PAGE gels (Thermo Fisher Scientific, \allowbreak\#NP0336BOX) and blotted on a 0.2 µm NC2 nitrocellulose membrane.
Membranes were cut at 100 kDa and blocked with 5\% BSA in TBS-Tween (0.1\%).
The following antibodies were used at 1:500 concentration and incubated overnight at 4 °C: phospho-VEGF Receptor 2 (Tyr1175) (Cell Signaling Technologies [CST], \#3770), VEGF Receptor 2 (CST, \#2479), phospho-p44/42 MAPK (Erk1/2) (CST, \#4377), p44/42 MAPK (Erk1/2) (CST, \#9102), phospho-Akt (Ser473) (CST, \#4060), Akt (CST, \#4691) and GAPDH (Novus Biologicals, \#NB300-221).
The following HRP-conjugated secondary antibodies were used at 1:5000 concentration: donkey anti-rabbit IgG (Abcam, \#ab16284) and donkey anti-mouse IgG (Abcam, \#ab98799).
Blots were developed using HRP substrate (Millipore, \#11556345) and were developed on an Amersham Imagequant 800. 
The phosphorylated version of each protein was detected on a different membrane than the non-phosphorylated protein.
The mean intensity for each band was measured using the same quantification area and the ratio of phosphorylated to non-phosphorylated protein was calculated.

\subsection{SARS-CoV-2 virus neutralization assay}
Experiments were performed by the Francis Crick Institute COVID Surveillance Unit following the protocol outlined in \citep{Shawe-Taylor2024-ko}.
Briefly, 10-point binder dose response series were generated by serially diluting each binder in duplicate in 20 mM sodium phosphate buffer before diluting further to achieve final testing concentrations of 1.7-11,200 ng/mL in 10\% fetal bovine serum (FBS).
With appropriate positive and negative controls, binder dose response series were then run through the standard live-virus neutralization assay against 2 variants of concern (VOC) and 2 variants of interest.
Duplicate assay plates were run, so each biological repeat contained 4 technical replicates.
Two biological repeats were run on separate days using different flasks of cells, vials of virus, and bottles of media.
Thus each plot of Supplementary Figure S11 consists of 160 independent data points. 
The data points were generated from 4 replicates of 40 independent titrations. {\EC} values were calculated using nonlinear regression with a 4-parameter dose response curve fit.

\subsection{Cryo-EM sample preparation, data collection and image processing}
The Spike ectodomain construct used in the cryo-EM experiments was based on the Wuhan SARS-CoV-2 isolate.
The protein (spanning residues 1-1208 from UniProt ID YP\_009724390) harbored point mutations K986P and V987P stabilizing the pre-fusion conformation, disrupted furin cleavage site, C-terminal T4 fibritin trimerization domain, and a hexa-histidine affinity tag \citep{Wrobel2020-mo}.
The protein was produced by expression in stably transformed Expi293F cells and purified by capture onto immobilized Ni affinity resin, followed by SEC, as previously described \citep{Wrobel2020-mo, Rosa2021-ei}.

Four µL freshly isolated trimeric SARS-CoV-2 Spike ectodomain (1.2 mg/mL), supplemented with 0.2 mg/mL GDM\_SC2RBD\_104, GDM\_SC2RBD\_50, GDM\_SC2RBD\_11, or GDM\_SC2RBD\_129 and 0.1\% n-octyl glucoside in 150 mM NaCl, 20 mM Tris-HCl, pH 8.0, was spotted onto fresh 400-mesh R1.2/1.3 C-flat holey carbon grids (Electron Microscopy Sciences product CF413-50-Au for 1 minute, under 100\% humidity at 20 °C, prior to blotting and plunge-freezing in liquid ethane-propane using Vitrobot Mark IV (Thermo Fisher Scientific).
Cryo-EM data were acquired on a Titan Krios G2 cryo-electron microscope equipped with a Falcon 4i direct electron detector (Thermo Fisher Scientific).
Selectris energy filter (Thermo Fisher Scientific) with a slit width of 10 eV was used for imaging complexes containing GDM\_SC2RBD\_11 and GDM\_SC2RBD\_129.
A total of 4500, 8342, 6728, and 8482 micrograph movies were recorded from grids containing GDM\_SC2RBD\_104, GDM\_SC2RBD\_50, GDM\_SC2RBD\_11, and GDM\_SC2RBD\_129, respectively.
Data collections proceeded with a defocus range set to -1.5 to -3.5 µm and a magnification corresponding to calibrated pixel size of 1.08 {\AA} (GDM\_SC2RBD\_104 and GDM\_SC2RBD\_50) or 0.95 {\AA} (GDM\_SC2RBD\_11 and GDM\_SC2RBD\_129) (\autoref{tab:cryoem_data_processing}).

\begin{figure}[H]
    \centering
    \begin{subfigure}[t]{0.62\textwidth}
        \centering
        \caption*{\textbf{A}}
        \includegraphics[]{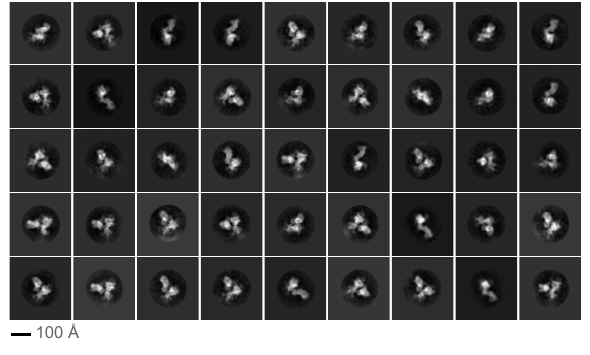}
        \phantomsubcaption\label{fig:sarscov2_cyroem_data_processing:1}
    \end{subfigure}
    \begin{subfigure}[t]{0.37\textwidth}
        \centering
        \caption*{\textbf{B}}
        \includegraphics[]{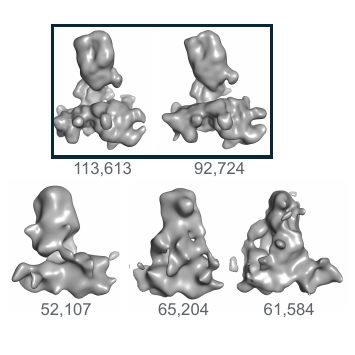}
        \phantomsubcaption\label{fig:sarscov2_cyroem_data_processing:2}
    \end{subfigure}
    \hfill
    \begin{subfigure}[t]{0.43\textwidth}
        \centering
        \caption*{\textbf{C}}
        \includegraphics[]{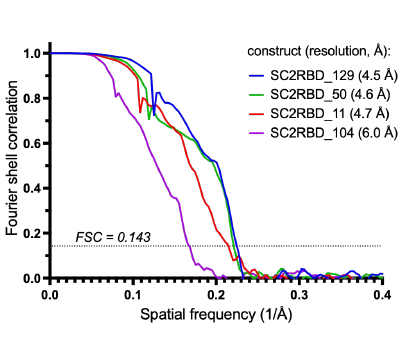}
        \phantomsubcaption\label{fig:sarscov2_cyroem_data_processing:3}
    \end{subfigure}
    \begin{subfigure}[t]{0.56\textwidth}
        \centering
        \caption*{\textbf{D}}
        \includegraphics[]{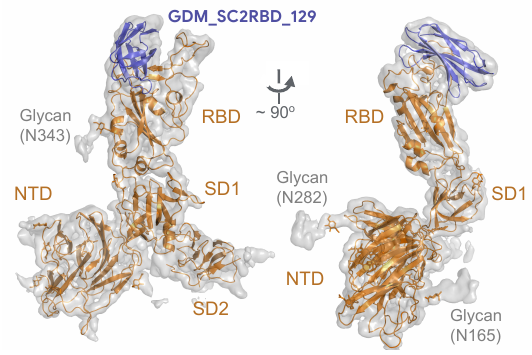}
        \phantomsubcaption\label{fig:sarscov2_cyroem_data_processing:4}
    \end{subfigure}
    \begin{subfigure}[t]{\textwidth}
        \centering
        \caption*{\textbf{E}}
        \includegraphics[]{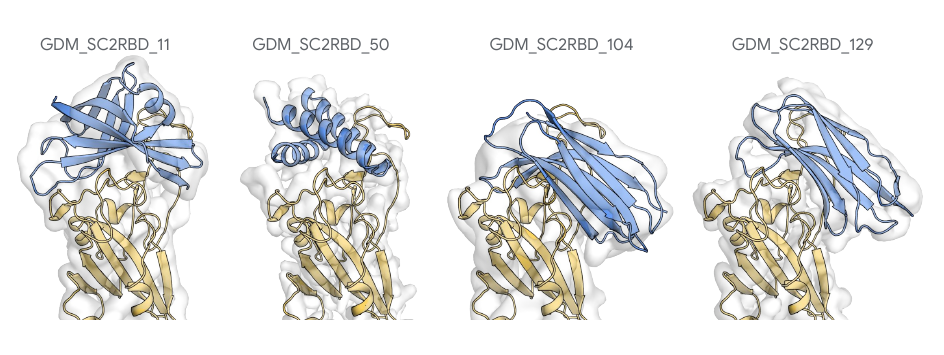}
        \phantomsubcaption\label{fig:sarscov2_cyroem_data_processing:5}
    \end{subfigure}
    \titledcaption{SARS-CoV-2 cryo-EM data processing.}{
        \textbf{(A)} 2D class averages of particle images from the SC2RBD\_129 dataset corresponding to dissociated spikes; the scale bar is 100 {\AA}.
        \textbf{(B)} The result of 3D classification of the same particles into 5 classes; the number of particles belonging to each class is indicated underneath.
        The two best 3D classes collectively comprising 206337 particles used for the final 3D reconstruction are boxed.
        \textbf{(C)} Half-map Fourier shell correlation (FSC) for each of the final reconstructions.
        Dotted line indicates the gold-standard cut-off at FSC of 0.143.
        \textbf{(D)} Final reconstruction of S1 in complex with GDM\_SC2RBD\_129 ligand in two orthogonal orientations. The cryo-EM map is shown as a transparent gray surface with docked S1 (from PDB ID 7ZBU, orange) and the ligand (blue) chains as cartoons.
        Locations of the ligand, individual S1 domains (RBD, NTD, SD1 and SD2) as well as select glycans (attached to Asn residues 165, 282, and 343) are indicated.
        \textbf{(E)} SARS-CoV-2 receptor binding domain (yellow) bound to the binder design (blue) was fitted together into the Cryo-EM density map (transparent gray surface).
    }
    \label{fig:sarscov2_cyroem_data_processing}
\end{figure}

1,674 EER frames recorded per micrograph movie were processed into 31 fractions, with an exposure dose of 1.04 e/Å2 (GDM\_SC2RBD\_104 and GDM\_SC2RBD\_50) or 1.25 e/Å2 (GDM\_SC2RBD\_11 and GDM\_SC2RBD\_129) per fraction.
The micrograph frames were aligned, summed and weighted as implemented in Relion-5.0beta \citep{Kimanius2021-yi, Zivanov2022-eq}, and contrast transfer function parameters were estimated using Gctf-v1.18 \citep{Zhang2016-zh}.
Reference-free 2D classification of an initial subset of particles picked using Gaussian blob function in Relion revealed 2D averages belonging to monomeric S1 protein, due to dissociation of the trimeric Spike.
Particles belonging to well-defined 2D classes were used to train Topaz \citep{Bepler2019-lp}, which was used to pick the entire datasets. The particles, extracted with 4-fold binning, were subjected to three rounds of 2D classification in Relion, using 400 classes in each round; the regularization parameter T was increased from 2 during the first round to 8 in the last round of 2D classification.
Particles contributing to well-defined 2D classes, re-extracted with 2-fold binning, were used to generate initial 3D models and subjected to 3D classification into 4-7 classes in Relion, with the regularization parameter T set to 8 (\autoref{tab:cryoem_data_processing},\autoref{fig:sarscov2_cyroem_data_processing}).
The best particle sets were used for 3D reconstruction, followed by Bayesian polishing \citep{Zivanov2022-eq}.
The final reconstructions were obtained using soft masks in conjunction with Blush regularization, as implemented in Relion-5.0beta \citep{Kimanius2024-ps}.
Resolution metrics reported in this work were according to the gold-standard Fourier shell correlation (FSC) 0.143 criterion \citep{Rosenthal2003-ux, Scheres2012-is}.
For illustration purposes, cryo-EM maps were locally filtered using EMReady \citep{He2023-me}.
Rigid body docking of S1 protein chain (from PDB ID 7ZBU) \citep{Seow2022-zs} and binder models into the final cryo-EM maps was done in Coot \citep{Emsley2004-ub}, and the figures were prepared using PyMOL Molecular Graphics System, Version 3.0 (Schrödinger, LLC).
Final cryo-EM maps will be deposited with the Electron Microscopy Data Bank (EMDB); the raw data will be available upon request.

\subsection{X-ray crystallography sample preparation, data processing and structure solving}
GDM\_VEGFA\_71 and VEGF-A were mixed in a molar ratio of 2.5:1, and incubated at room temperature for 1 hour with shaking at 1000 rpm. The GDM\_VEGFA\_71/VEGF-A complex was purified by SEC using a Superdex 200 Increase 10/300 GL column (Cytiva), equilibrated with 20 mM Tris pH 7.5, 150 mM NaCl, and verified by SDS-PAGE. The GDM\_VEGFA\_71/VEGF-A complex was concentrated to 12 mg/mL using a 10 kDa MWCO concentrator (Vivaspin). Crystallisation was performed using a Mosquito crystallization robot (SPT Labtech) by sitting-drop vapor diffusion (50 nL complex + 50 nL crystallization solution) in 3-well crystallization plates (SWISSCI) containing 25 µL of crystallization solutions in each reservoir. Crystals of the protein complex grew within two weeks at 20 °C in mother liquor containing 0.1 M phosphate/citrate pH 4.2 and 40\% v/v ethanol.
Crystals were harvested with 10 µm Micromount loops (MiTeGen) and snap-frozen in liquid nitrogen prior to data collection. X-ray data were collected from a single crystal at 100 K on the I04 beamline at Diamond Light Source (Harwell, UK) with a wavelength of 0.9537 {\AA}. 
All data were automatically processed by xia2 \citep{Gildea2022-bl}. Initial phases for the GDM\_VEGFA\_71/VEGF-A complex were obtained by maximum-likelihood molecular replacement using Phaser (version 2.8.3) \citep{McCoy2007-sl} from the CCP4 Suite (version 9.0.002) \citep{Agirre2023-yd} using the AF3-predicted structure as a search model. The structure solution was subjected to repetitive rounds of restrained refinement using Refmac5 (version 5.8.0430)~\citep{Murshudov2011-gr} and interactive manual building in COOT (version 0.9.8.95)~\citep{Emsley2010-fo}. NCS and Jelly Body restraints were also used throughout the refinement. The final structure quality at 2.56 Å was assessed using Molprobity (version 0.9.8.95)~\citep{Williams2018-cb}.
Data collection and refinement statistics are provided in \autoref{tab:vegfa_xtal_stats}.

\section{Iterative development and \emph{in silico} benchmarking of AlphaProteo}
\label{sec:in_silico_benchmark}
During development of AlphaProteo, we trained two versions of the generative model (\autoref{fig:overview:pipeline_schematic}), referred to here as "v1" and "v2".
To evaluate these models, we used two different \emph{in silico} benchmarks, each consisting of a set of targets along with a definition of success rate, or the fraction of designs satisfying certain computational success criteria.
We compared the v1 and v2 generative models to the current best binder-design method RFdiffusion~\citep{Watson2023-fy}.

\begin{figure}[H]
    \centering
    \vspace{-0.2cm}
    \begin{subfigure}[t]{0.49\textwidth}
        \centering
        \caption*{\textbf{A}}
        \includegraphics[]{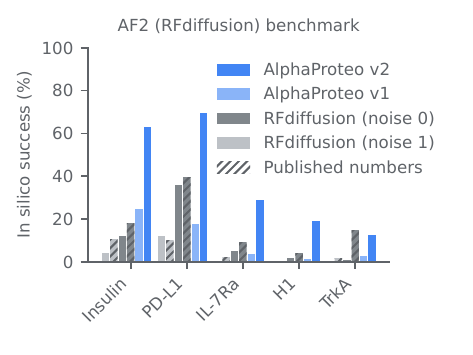}
        \setcounter{subfigure}{0}
        \phantomsubcaption\label{fig:in_silico_results:af2_ig_benchmark}
    \end{subfigure}
    \hfill
    \begin{subfigure}[t]{0.49\textwidth}
        \centering
        \caption*{\textbf{C}}
        \includegraphics[]{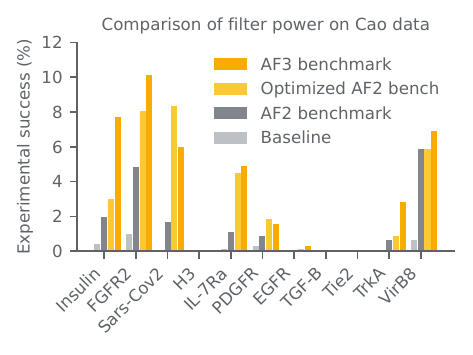}
        \setcounter{subfigure}{2}
        \phantomsubcaption\label{fig:in_silico_results:cao_optimized_filters}
    \end{subfigure}
    \vspace{-1.5cm}
    \vfill
    \begin{subfigure}[t]{0.98\textwidth}
        \centering
        \caption*{\textbf{B}} 
        \includegraphics[]{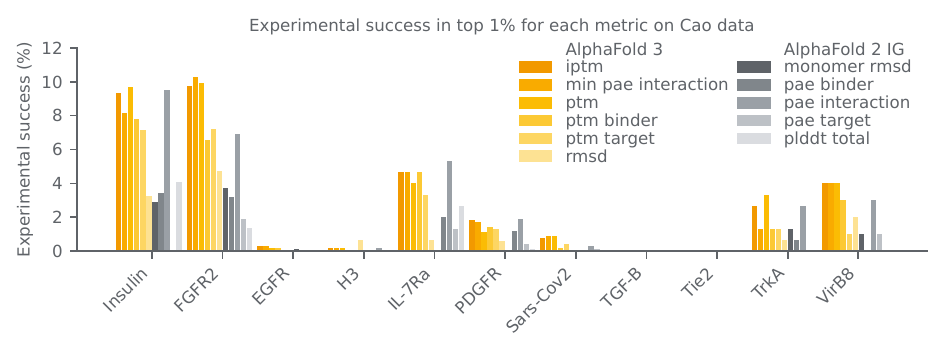}
        \setcounter{subfigure}{1}
        \phantomcaption\label{fig:in_silico_results:cao_success_by_metric}
    \end{subfigure}
    \vspace{-1.1cm}
    \vfill
    \begin{subfigure}[t]{0.98\textwidth}
        \centering
        \caption*{\textbf{D}}
        \includegraphics[]{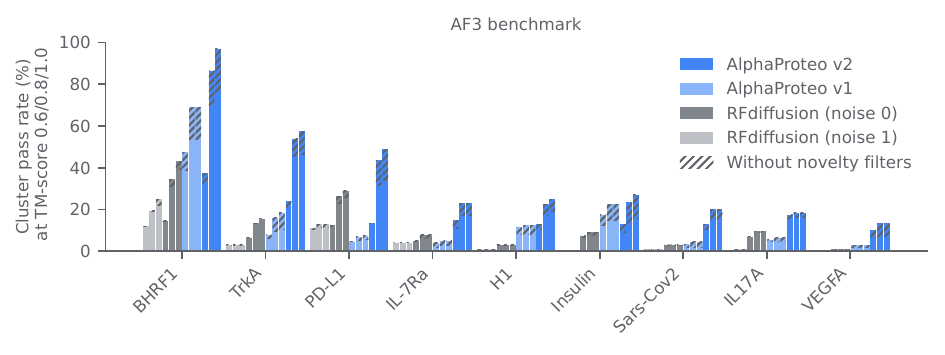}
        \setcounter{subfigure}{3}
        \phantomcaption\label{fig:in_silico_results:af3_binder_benchmark}
    \end{subfigure}
    \vspace{-0.5cm}
    \titledcaption{\emph{In-silico} performance of AlphaProteo and development of an AF3-based binder design benchmark.}{
        See \autoref{sec:in_silico_benchmark} for full details.
        \textbf{(A)} \emph{In silico} success rates of AlphaProteo and RFdiffusion under the "AF2 (RFdiffusion) benchmark", which consists of the AF2 initial guess prediction method, scoring thresholds, and design targets described in \citep{Watson2023-fy}.
        For RFdiffusion, both the published values and our own reproduction of its performance are shown.
        \textbf{(B)} Retrospective experimental success rate of designs from \citep{Cao2022-vw} with the top 1\% values of each AF2- or AF3-derived metric. This identifies the metrics that individually have the strongest predictive value for experimental success.
        \textbf{(C)} Retrospective success rate of designs from \citep{Cao2022-vw} after filtering by different definitions of \emph{in silico} success: "Baseline": fraction of successful binders in the unfiltered data from \citep{Cao2022-vw}; "AF2 benchmark": metrics and filtering thresholds used in \citep{Watson2023-fy} and \textbf{(A)}; "Optimized AF2 benchmark": optimized thresholds on the same metrics used in the AF2 benchmark; "AF3 benchmark": optimized thresholds on a small set of the most predictive AF3 metrics from \textbf{(B)}. The "AF3 benchmark" filtering criteria enrich most strongly for experimental success.
        \textbf{(D)} \emph{In silico} success rates of AlphaProteo and RFdiffusion under the "AF3 benchmark", consisting of both the targets in this work and the previous AF2 benchmark targets, along with optimized AF3 metrics and thresholds as shown in \textbf{(C)}.
        Clustered bars of the same color show diversity-adjusted success rates via pairwise TM-score clustering at different thresholds (0.6/0.8/1.0). Hatched bars show the reduction in success rate after excluding designs with sequence bit-score > 50 in pHMMER search against the Uniref50 dataset.
    }
    \label{fig:in_silico_results}
\end{figure}

First, we used an existing binder design benchmark based on AlphaFold 2 (AF2), where a designed binder against each of 5 target proteins is considered a success if its AF2 prediction has interchain predicted aligned error < 10, binder-aligned binder RMSD < 1 Å, and pLDDT > 80 (see detailed methods below).
These criteria were shown to be highly predictive of experimental binding success \citep{Bennett2023-cn}. On this benchmark, the v2 generative model has higher success rates than RFdiffusion on 4 out of 5 targets (\autoref{fig:in_silico_results:af2_ig_benchmark}).
The v1 model outperformed RFdiffusion on 4 of 5 targets when RFdiffusion is run at noise level 1 but underperformed RFdiffusion at noise level 0.

Given that AlphaFold 3 (AF3) is more accurate than AF2 on protein complex prediction, we developed a second benchmark based on AF3~\citep{Abramson2024-nx}.
On a set of 9 targets, we considered a design a success if its minimum interchain predicted aligned error < 1.5, predicted TM-score > 0.8, and complex RMSD < 2.5.
We found these optimized criteria to be a better proxy of experimental success than AF2 on a published \emph{de novo} binder dataset (\autoref{fig:in_silico_results:cao_success_by_metric}, \autoref{fig:in_silico_results:cao_optimized_filters}) \citep{Cao2022-vw}.
On this benchmark, the v2 model had higher \emph{in silico} success rates than RFdiffusion on all targets and the v1 model outperforms both RFdiffusion variants on six of nine targets (\autoref{fig:in_silico_results:cao_optimized_filters}).
These conclusions do not change when success rate is adjusted to account for diversity (pairwise TM-score clustering at various thresholds) or novelty (pHMMER bit-score < 50 against Uniref50) (\autoref{sec:methods}, \autoref{fig:in_silico_results:af3_binder_benchmark}).
Taken together, these results show that the \emph{in silico} performance of AlphaProteo is at or above the SoTA, consistent with our experimental results.

We experimentally tested a design system containing the v1 model against all 7 targets, a v1-based system with an improved filter against SC2RBD and PD-L1, and a v2-based system on PD-L1 and TrkA (\autoref{tab:internal_model_comparison}).
Both improving the filter and the model resulted in increased experimental success rates.
For simplicity, the results in \autoref{tab:experimental_results} are pooled over all tested designs for each target.
Importantly, all designs tested in this work were generated in a "zero-shot" manner, without using any known binder as a starting point.

\subsection{AF2-based benchmark}
We followed published procedures \citep{Watson2023-fy} to run a previously proposed AF2 benchmark \citep{Bennett2023-cn} on AlphaProteo designs.
This includes generating designs for IL-7RA, PD-L1, TrkA, Insulin receptor, and hemagglutinin H1 using published input specifications (\autoref{tab:binder_specs}) \citep{Watson2023-fy}, inputting them to the "AlphaFold2 initial guess" script (AF2ig) \citep{Watson2023-fy, Bennett2023-cn}, and computing the fraction of successful designs, defined as those with interchain AF2 pAE < 10, binder-aligned binder RMSD < 1 Å, and plDDT > 80.
All mentions of "AlphaFold 2" in the context of benchmarks refer to the AF2ig method; we did not run unmodified AF2 for any analyses in this work.
We generated 200 designs per target per model.
Additionally, we downloaded RFdiffusion\footnote{\url{https://github.com/RosettaCommons/RFdiffusion}} and ran it on the benchmark using both noise=0 and noise=1 settings to ensure we could reproduce the published performance of RFdiffusion.
As in \citep{Watson2023-fy}, we redesigned sequences with ProteinMPNN \citep{Dauparas2022-fc} with low sampling temperature 0.0001. We present RFdiffusion success rates from both the original publication and from our reproduction.

\subsection{AF3-based benchmark}
\label{sec:af3bench}
To create an AF3-based \emph{in silico} benchmark for binder design, we looked for metrics and thresholds that most enrich for experimental success across 640,000 previously characterized \emph{de novo} binder designs against 11 targets~\citep{Cao2022-vw}.
We used this dataset because it was not filtered on any AF2- or AF3-based metrics prior to experimental testing and therefore any filters we derive from it have the best chance of generalizing to future design methods.
We predicted the structure of each binder-target complex using AF3 while inputting the structure of the target as a template and using only a single sequence (no multiple sequence alignment).
We selected the best out of five diffusion head samples using a ranking confidence of $(0.8\;\text{iptm} + 0.2\;\text{ptm})$, the individual components of which are described below.
We computed the retrospective success rate (fraction of designs with observed binding at 4000 nM target) among the top 1\% of designs according to a panel of AF2- and AF3-based metrics (\autoref{fig:in_silico_results:cao_success_by_metric}):
\newpage 
\textbf{AF3} (see Supplementary Information of \citep{Abramson2024-nx} for details)
\begin{enumerate}
    \item \textbf{ptm}: prediction aligned error (PAE) matrix reduction, maximum average error across aligning on individual residues.
    \item \textbf{ptm binder / ptm target}: intra-chain reduction of the PAE matrix, aligning on binder / target chain residues and considering errors on the same chain.
    \item \textbf{iptm}: interchain reduction of the PAE matrix, taking into account only those PAE entries for TM computation that are not on the chain that is being aligned on.
    \item \textbf{min pae interaction}: minimum value across all interchain terms in the PAE matrix.
    \item \textbf{rmsd}: root mean squared error between the designed and predicted complex structures.
\end{enumerate}

\textbf{AF2 initial guess}
\begin{enumerate}
    \item \textbf{pae binder / pae target}: average of the PAE matrix when only considering the binder / target chain.
    \item \textbf{pae interaction}: average PAE of the interchain residues.
    \item \textbf{plddt total}: average plddt over the predicted complex structure.
    \item \textbf{monomer rmsd}: root mean squared error between the binder design and prediction when aligning on the binder chain.
\end{enumerate}

We developed a new definition of \emph{in silico} success by performing a combinatorial sweep over the following grid of filtering metric thresholds (start, stop, step):

\begin{center}
    \begin{tabular}{llll}
         \textbf{AF3} & & \textbf{AF2 initial guess}\\
         min pae interaction & (0, 7, 0.5) & af2 monomer rmsd & (0, 3, 0.5)\\
         ptm binder & (0, 1, 0.05) & pae interaction & (0, 11, 0.5)\\
         rmsd & (0, 3, 0.5) & plddt & (60, 95, 5)\\
    \end{tabular}
\end{center}

For each target, we ranked the different filter settings according to the binding success rate among passing examples from the data collected by \citet{Cao2022-vw}.
We optimized the average per-target rank subject to the constraint that at least 10 designs have to pass filters.
We chose to aggregate performance across targets by rank rather than by pass rates due to large variability in the latter.
This yielded the following "optimized" filtering thresholds for both AF3 and AF2:

\begin{center}
    \begin{tabular}{llll}
         \textbf{AF3} & & \textbf{AF2 initial guess}\\
         min pae interaction & $< 1.5$ & af2 monomer rmsd & $< 1.5$\\
         ptm binder & $> 0.8$ & pae interaction & $< 7.0$\\
         rmsd & $< 2.5$ & plddt & $> 90$\\
    \end{tabular}
\end{center}

Per-target retrospective success rates based on these filters are shown in \autoref{fig:in_silico_results:cao_optimized_filters}. 
As the AF3 optimized filters slightly outperform the AF2 optimized ones across all targets, we used the AF3 filters to define \emph{in silico} success for a new benchmark.
We then computed success rates (using the optimized AF3 thresholds) on designs from AlphaProteo models v1 and v2, as well as our local installation of RFdiffusion.
As targets, we selected the original RFdiffusion design targets as well as new targets that we addressed experimentally in this work (\autoref{tab:binder_specs}).
To account for structural diversity, we used TM-align to compute pairwise TM-scores separately for designs sampled for each target and from each model, and we used a greedy algorithm to cluster these designs at a given TM-score threshold.
To account for novelty, we searched each design sequence against Uniref50 using Jackhmmer~\citep{Johnson2010-bp} and considered it novel if its maximum bit-score is less than 50.

\section{\textit{In silico} screening of PDB targets}
\label{sec:in_silico_target_selection}

\begin{figure}[H]
    \centering
    \includegraphics[]{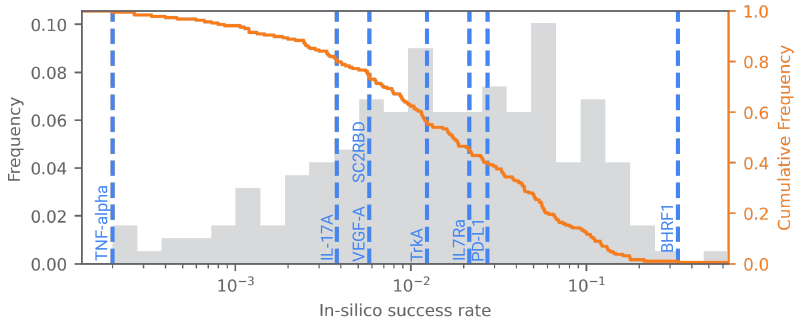}
    \titledcaption{Distribution of \emph{in silico} success rates.} {Histogram (gray) and complementary cumulative density (orange line) of \textit{in silico} success rates for AlphaProteo binder design against 200 randomly sampled target proteins from the PDB (\autoref{sec:in_silico_target_selection}). The 7 targets for which we successfully obtained binders (labeled blue dotted lines) cover a broad range of \textit{in silico} success rates. TNF$\alpha$, where we failed to obtain binders, is among the most challenging \textit{in silico} targets, while IL-17A, where we succeeded experimentally, is more difficult than $80\%$ of the \textit{in silico} targets.}
    \label{fig:in_silico_target_histogram}
\end{figure}

In order to estimate and compare the difficulty of potential future binder design problems to the 8 targets we experimentally evaluated, we computed \textit{in silico} success rates for a random subset of target proteins from the PDB. Starting from 45k clusters derived at $40\%$ sequence homology~\citep{Abramson2024-nx}, we filtered for PDB entries 
containing up to 5 protein chains and ranging from 30 to 400 residues in length, to exclude those too large to process efficiently. After removing singleton clusters, we sampled 1 protein representative from 200 randomly selected clusters (out of the remaining 6000 final clusters). For each of the 200 proteins, we sampled 1 chain at random to serve as the target, and sampled 3x distinct regions on the protein surface to serve as binding hotspots. Finally, we generated 5000 binders for each target:hotspot combination and computed their \textit{in silico} success rate, or fraction of designs predicted to bind according to the Alphafold 3 benchmark criteria (\autoref{sec:af3bench}). Overall, 600 target epitopes were tested; these were pooled by target, for a total of 200 \textit{in silico} success rates plotted in \autoref{fig:in_silico_target_histogram}.

\section{Comparison to other design methods}
\label{sec:comparison_to_other_methods}

\subsection{Comparison of experimental success rates to RFdiffusion}
Published experimental success rates of RFdiffusion were based on 96-well biolayer interferometry (BLI) measurements~\citep{Watson2023-fy}, which are impractical for us to perform on our own designs.
Therefore, to quantitatively compare success rates, we downloaded sequences of the published RFdiffusion designs\footnote{\url{https://figshare.com/s/439fdd59488215753bc3}} and synthesized and tested them by yeast display (\autoref{sec:ysd}) alongside our own designs.
Our results matched the published success rates for PD-L1 but were 2-fold lower for IL-7RA (16\% versus 33\%) and 0\% for TrkA (versus 6\%) (\autoref{tab:internal_model_comparison}).
This is potentially due to differences between the yeast surface environment versus purified proteins, as well as our use of 10- to 20-fold lower target concentrations (0.5-1 µM versus 10 µM in \citep{Watson2023-fy}), which may exclude lower-affinity hits.

\subsection{Comparison of binding affinity (\KD) to other methods}
\begin{figure}[H]
    \centering
    \includegraphics[]{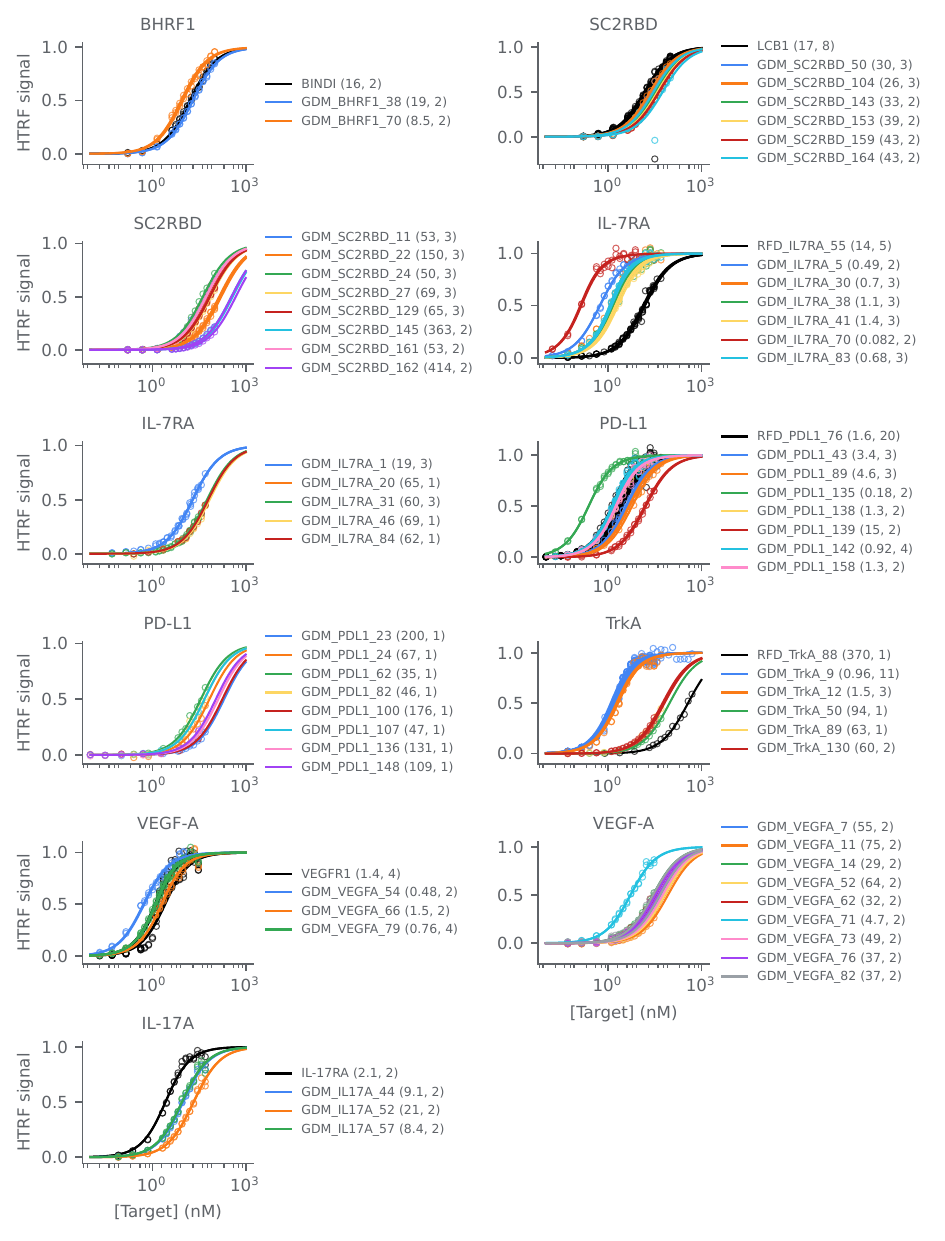}
    \titledcaption{Equilibrium saturation binding by homogeneous time-resolved fluorescence (HTRF).}{
        HTRF equilibrium saturation binding affinity measurement data for all designs that expressed and displayed observable binding.
        Control binders from the literature are shown in black.
        The binder design was held at a fixed concentration of either 0.1 nM or 1 nM and the target protein concentration was titrated. Parentheses indicate the fitted {\KD} value from a generalized (square-root form) 1:1 binding equation (\autoref{sec:htrf}) and the number of replicates for each design.
        Note that the x-axis is in log-scale to highlight order-of-magnitude {\KD} differences, but this gives the appearance that data is less saturated than it is. We report the inter-replicate variation as well as the fitting uncertainty in (\autoref{tab:sequences}).
    }
    \label{fig:htrf_all}
\end{figure}

\begin{figure}[H]
    \centering
    \includegraphics[]{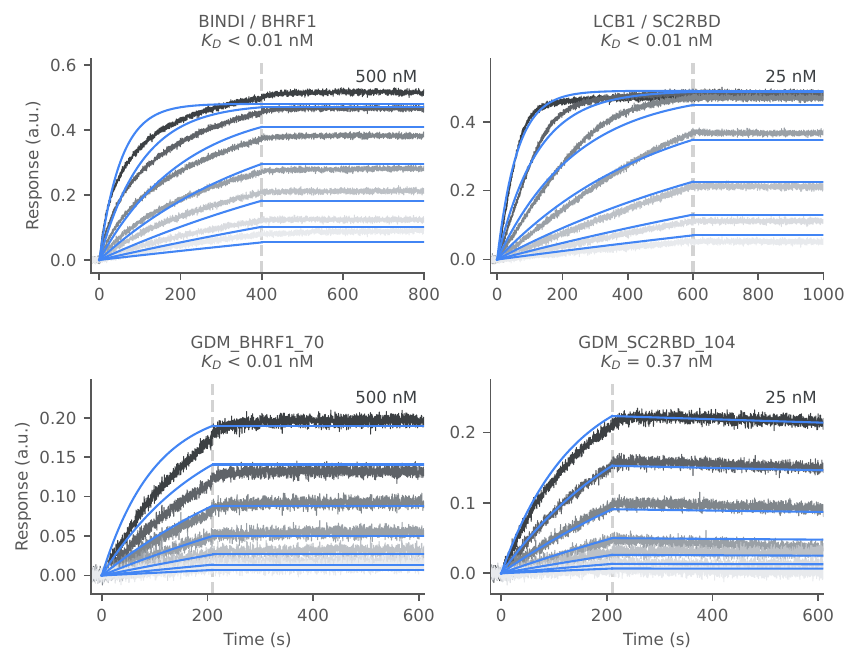}
    \titledcaption{Binding kinetics by biolayer interferometry (BLI).}{
        BLI kinetic data (gray gradient), fitted 1:1 binding models with shared global {$\mathrm{R_{max}}$} (blue lines), and resulting {\KD} estimates for selected controls and designs (\autoref{sec:bli}). Analyte concentrations are in a 2-fold serial dilution from the maximum concentration indicated. "$K_D < 0.01$" indicates a fitted {\KD} value that is too low to interpret reliably due to lack of apparent dissociation.
        
    }
    \label{fig:bli}
\end{figure}

We included the best designed binders from the literature for BHRF1 \citep{Procko2014-az}, SC2RBD \citep{Cao2020-cy}, IL-7RA, PD-L1, and TrkA \citep{Watson2023-fy} as controls in our HTRF {\KD} measurements (\autoref{tab:experimental_results}, \autoref{tab:internal_model_comparison}, \autoref{tab:successful_yeast_hits}; \autoref{fig:htrf_all}).
For the TrkA and IL-7RA binders we successfully reproduced the literature values, obtaining 10\% higher and 2-fold lower {\KD} values, respectively, than previously reported. For PD-L1, because we also screened the original set of RFdiffusion designs, we found a design with 1000-fold better {\KD} (RFD\_PDL1\_76, KD=1.6 nM) than the published "best" design (RFD\_PDL1\_77, \KD=1.4 µM) \citep{Watson2023-fy}.
Therefore, we compared our results to the higher-affinity PD-L1 binder (\autoref{tab:experimental_results}, \autoref{tab:internal_model_comparison}).
For the BHRF1 and SC2RBD control binders ("BINDI" and "LCB1", respectively), we obtained {\KD} values $\sim$10-fold higher than what was reported previously from biolayer interferometry (BLI) experiments \citep{Procko2014-az, Cao2020-cy}.
A possible explanation is that proteins are in solution and in equilibrium in HTRF while in BLI, one species is immobilized on a 2-dimensional surface.
This may allow the mobile species to rebind another molecule of the immobilized species without dissociating from the surface, increasing the apparent binding affinity relative to the solution-phase \citep{Daniel2013-oa}.
We verified that BINDI and LCB1, as well as AlphaProteo's best binders for these targets, have {\KD}<1 nM in our own BLI experiments (\autoref{fig:bli}).
As the exact {\KD} values from BLI are not quantitative in this regime, we compare these binders on the basis of our HTRF measurements (\autoref{tab:experimental_results}).

\section*{Supplementary figures}
\addcontentsline{toc}{section}{Supplementary figures}


\begin{figure}[H]
    \centering
    \begin{subfigure}[t]{\textwidth}
        \centering
        \caption*{\textbf{A}}
        \includegraphics[]{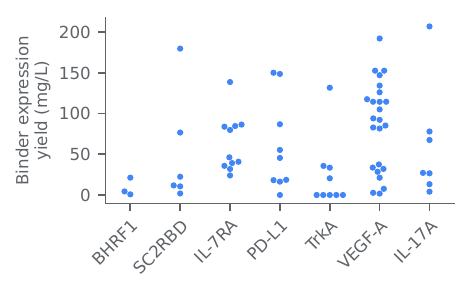}
        \phantomsubcaption\label{fig:expression_yield_and_sec:expression_yield}
    \end{subfigure}
    \hfill
    \begin{subfigure}[t]{\textwidth}
        \centering
        \caption*{\textbf{B}}
        \includegraphics[]{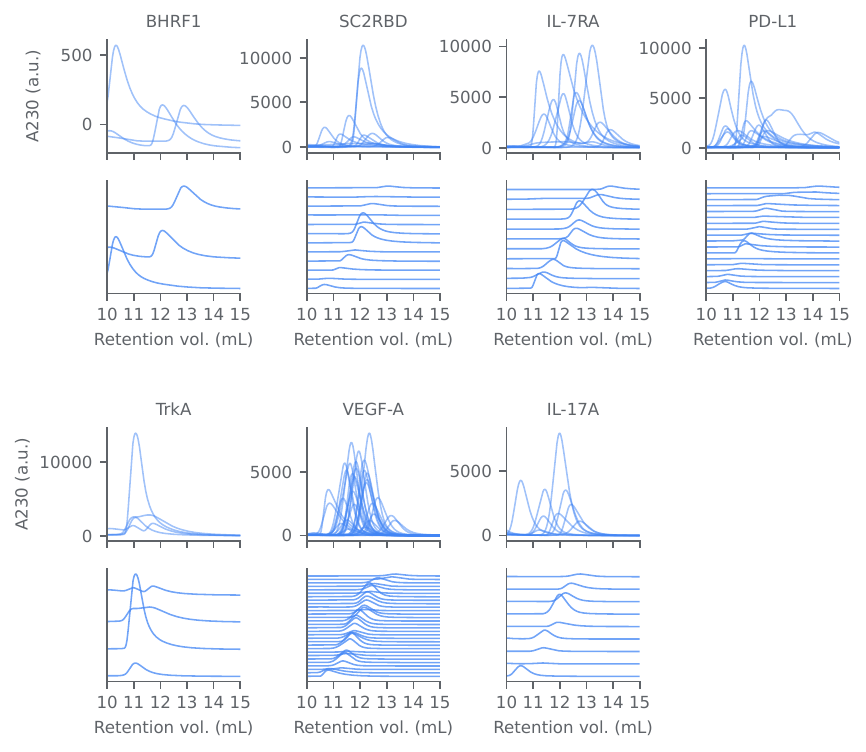}
        \phantomsubcaption\label{fig:expression_yield_and_sec:sec}
    \end{subfigure}
    \titledcaption{ Expression yield and size-exclusion chromatography of binder hits.}{
        \textbf{(A)} Protein yield from 10~mL \emph{E. coli} expression of yeast display binding hits, showing that most binders are highly expressed.
        \textbf{(B)} Size-exclusion chromatography of binding hits, showing that most binders are monodisperse and likely monomeric.
    }
    \label{fig:expression_yield_and_sec}
\end{figure}

\begin{figure}[H]
    \centering
    \includegraphics[]{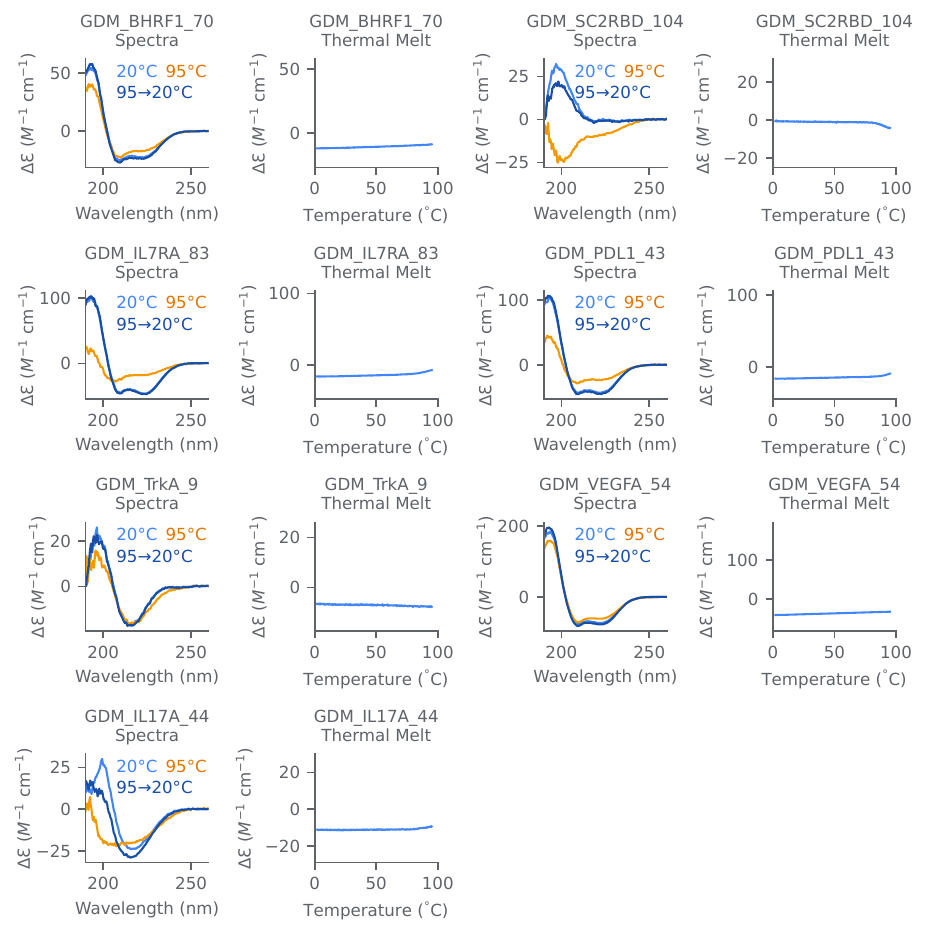}
    \titledcaption{Circular dichroism spectra and thermal melts.}{
        Circular dichroism spectra before (20 ºC) and after melting (95 ºC and 95 $\rightarrow$ 20 ºC).
        For most designs tested, spectra show expected secondary structure and refolding.
    }
    \label{fig:cd_spectra}
\end{figure}

\begin{figure}[H]
    \centering
    \includegraphics[]{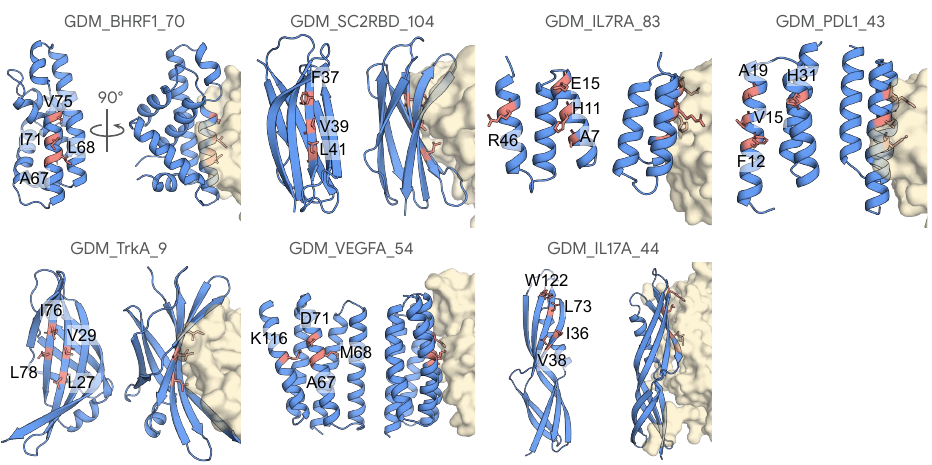}
    \titledcaption{Structural locations of interface mutations.}{
        Examples of interface residues (red) that were mutated 
        for one representative binder (blue) per target (pale yellow).
    }
    \label{fig:mutations_structures}
\end{figure}

\begin{figure}[H]
    \centering
    \includegraphics[]{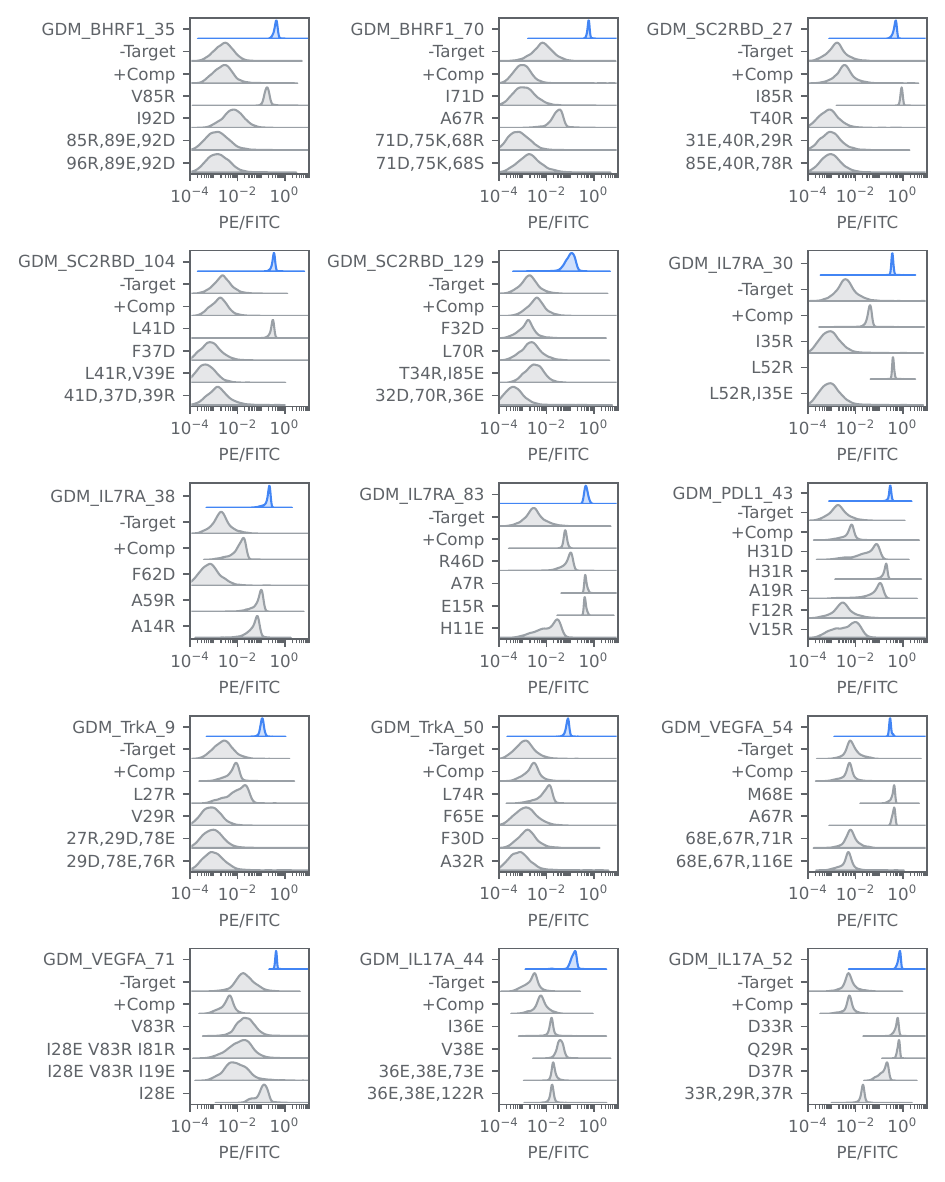}
    \titledcaption{Interface mutation and competition experiments on the best binders.}{
        Distributions of PE (binding) normalized to FITC (expression) from flow cytometry yeast display for interface mutants of selected binders as well as competitive inhibition by a previous known binder (\autoref{sec:mutants_and_competition}).
        In most cases, mutating the interface residue or adding a competitive binder decreases the binding signal, confirming that the design binds via the intended interface residues to the intended target site.
    }
    \label{fig:supp_mutants}
\end{figure}

\begin{figure}[H]
    \centering
    \begin{subfigure}[t]{\textwidth}
        \centering
        \caption*{\textbf{A}}
        \includegraphics[]{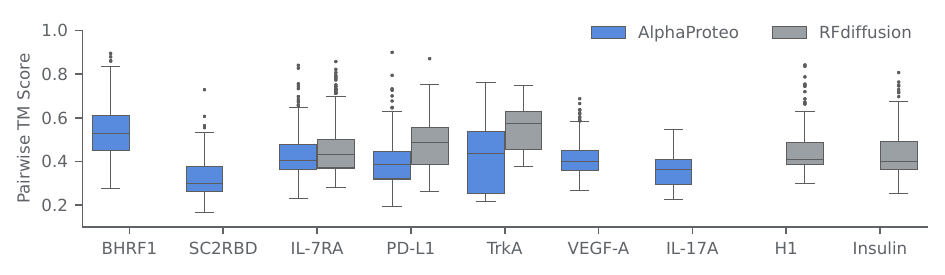}
        \phantomsubcaption\label{fig:diversity:pairwise_tm}
    \end{subfigure}
    \hfill
    \begin{subfigure}[t]{\textwidth}
        \centering
        \caption*{\textbf{B}}
        \includegraphics[]{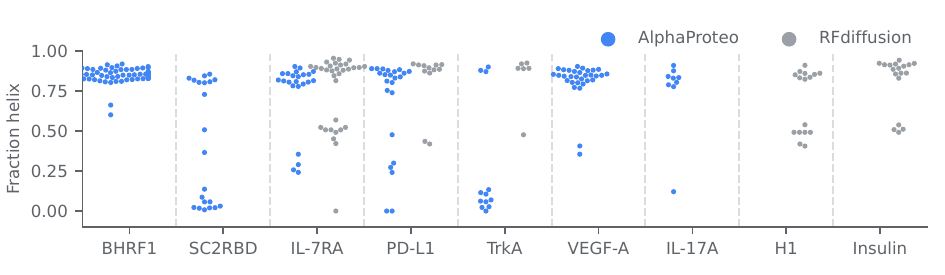}
        \phantomsubcaption\label{fig:diversity:secondary_structure}
    \end{subfigure}
    \titledcaption{Diversity of experimentally tested binder designs.}{
        \textbf{(A)} Pairwise TM score among designs from AlphaProteo and RFdiffusion with experimentally confirmed binding activity, including additional RFdiffusion targets H1 and Insulin \citep{Watson2023-fy}.
        Experimental success of RFdiffusion designs in this plot is based on the published results rather than our yeast display measurements, as our assay did not observe binding for any TrkA designs (\autoref{sec:comparison_to_other_methods}).
        \textbf{(B)} Fraction of binder residues annotated by DSSP as helix. The fraction of loop-annotated residues is low and relatively constant for all designs, so the fraction of beta sheets is 1 - fraction helix.
    }
    \label{fig:diversity}
\end{figure}

\begin{figure}[H]
    \centering
    \includegraphics[]{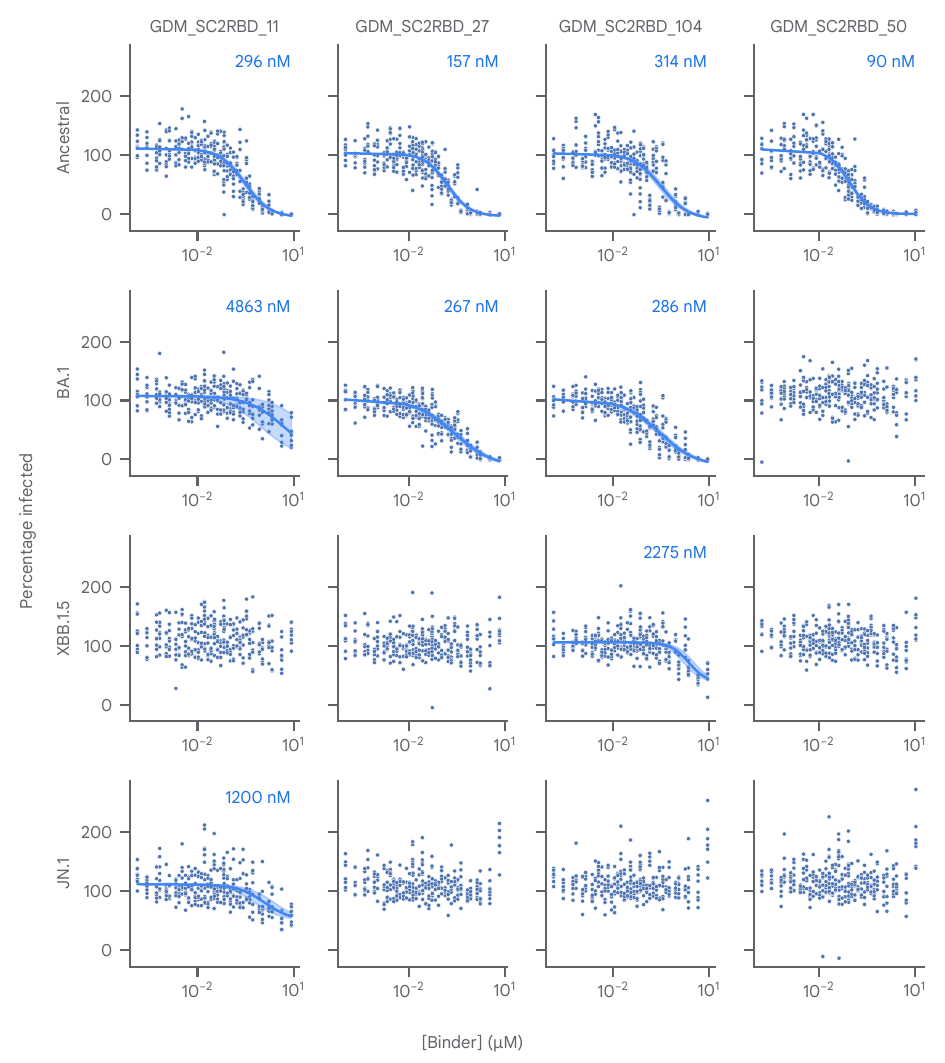}
    \titledcaption{SARS-CoV-2 neutralization assay for four selected binders over four variants of interest.}{
        SARS-CoV-2 virus neutralization assay was performed in Vero cells by the Francis Crick Institute COVID Surveillance Unit following the protocol outlined in \citep{Shawe-Taylor2024-ko}.
        Each plot consists of 160 independent data points (4 technical replicates, 2 biological replicates, 40 independent titrations).
        {\EC} values were calculated using nonlinear regression with a 4-parameter dose response curve fit.
        Fits are shown only when standard error on {\EC} was within one order of magnitude and the percentage infected is reduced to at least 80\%.
        Standard errors at each dilution are shown as shaded areas. All four binders tested successfully neutralize the ancestral variant of the SARS-CoV-2 virus.
    }
    \label{fig:sarscov2_neutralization}
\end{figure}

\clearpage
\section*{Supplementary tables}
\addcontentsline{toc}{section}{Supplementary tables}

\begin{table}[H]
  \small
  \centering
  \renewcommand\arraystretch{1.5}
  \begin{threeparttable}
    \begin{tabular}{@{}ll*{6}{>{\raggedright\arraybackslash}p{0.16\linewidth}}@{}}
      \toprule
      Design target & PDB ID & Target chain and residue numbers & Target hotspot residues & Natural binding partner & Binder length range (for benchmarks)\\
      \midrule
                BHRF1      & 2wh6  & A2-158                         & A65, A74, A77, A82, A85, A93  & BH3 helix             & 80-120                 \\
                SC2RBD     & 6m0j  & E333-526                       & E485, E489, E494, E500, E505 & ACE2 receptor        & 80-120                 \\
                IL-7RA     & 3di3  & B17-209                         & B58, B80, B139             & IL-7                  & 50-120                 \\
                PD-L1      & 5o45  & A17-132                         & A56, A115, A123           & PD-1                  & 50-120                 \\
                TrkA       & 1www  & X282-382                       & X294, X296, X333           & Nerve growth factor   & 50-120                 \\
                Insulin    & 4zxb  & E6-155                          & E64, E88, E96             & Insulin receptor     & 40-120                 \\
                H1         & 5vli  & A1-50, A76-80, A107-111, A258-322, B1-68, B80-170       & B21, B45, B52             & None                  & 40-120                 \\
                VEGF-A     & 1bj1  & V14-107, W14-107               & W81, W83, W91             & VEGFR1, VEGFR2        & 50-140                 \\
                IL-17A     & 4hsa  & A17-131, B19-127               & A94, A116, B67             & IL-17 receptor alpha & 50-140                 \\
                TNF$\alpha$  & 1tnf  & A12-157, B12-157, C12-157      & A113, C73                  & None  & 50-120                 \\

      \bottomrule
    \end{tabular}
    \titledcaption{Binder design problem specifications for \emph{in silico} benchmarking and experimental testing.}{
        Input target structures and hotspot residues for the design targets addressed here.
        For IL-7RA, PD-L1, TrkA, we used the same definitions as RFdiffusion \citep{Watson2023-fy}.
        For other targets, we used the highest resolution crystal structure available, and chose hotspot residues at the binding site of a natural interaction partner.
        When making designs for experimental validation, we used a binder length range of 50-140 amino acids.
    }
    \label{tab:binder_specs}
  \end{threeparttable}
\end{table}

\begin{table}[H]
    \small
    \centering
    \renewcommand\arraystretch{1.8}
    \begin{threeparttable}
        \begin{tabular}{p{0.15\linewidth}l*{8}{>{\centering\arraybackslash}p{0.08\linewidth}}@{}}
            \toprule
             & BHRF1 & SC2RBD & IL-7RA & PD-L1 & TrkA & IL-17A & VEGF-A & TNF$\alpha$\\
             \midrule
             & \multicolumn{7}{c}{\makecell[c]{\textbf{Success} (\%)\\\footnotesize{(higher is better)}}}\\
             AlphaProteo v1 & \mc{88}{94} & \makecell[t]{\mc{8.5}{94} \!\vrule\! \mc{8.5\tnote{$\dagger$}}{47}} & \mc{24.5}{94} & \mc{9.6}{94} & \mc{9.6}{94} & \mc{14.3}{63} & \mc{33}{94} & \mc{0}{30} \\
             \makecell[lt]{AlphaProteo v1\\(improved filters)} & \nan & \mc{29}{31} & \nan & \mc{16.1}{31} & \nan & \nan & \nan & \nan \\
             AlphaProteo v2 & \nan & \nan & \nan & \mc{26.5}{34} & \mc{8.1}{37} & \nan & \nan & \mc{0}{24} \\
             \makecell[lt]{RFdiffusion\\(our measurement)} & \nan & \nan & \mc{16.8}{95} & \mc{12.6}{95} & \mc{0.0}{95} & \nan & \nan & \nan\\
             \makecell[lt]{RFdiffusion\\(published)} & \nan & \nan & \mc{33.7}{95} & \mc{12.6}{95} & \mc{6.3}{95} & \nan & \nan & \nan \\
             \midrule
             & \multicolumn{7}{c}{\makecell[c]{\textbf{Binding $\mathrm{K_D} \mathrm{(nM)}$}\\\footnotesize{(lower is better)}}}\\
             AlphaProteo v1 & \mc{8.5}{94} & \makecell[t]{\mc{30}{94} \!\vrule\! \mc{26\tnote{$\dagger$}}{47}} & \mc{0.08}{94} & \mc{3.4}{94} & \mc{0.96}{94} & \mc{8.4}{63} & \mc{0.48}{94} & \nan\\
             \makecell[lt]{AlphaProteo v1\\(improved filters)} & \nan & \mc{33}{31} & \nan & \mc{47}{31} & \nan & \nan & \nan & \nan\\
             AlphaProteo v2 & \nan & \nan & \nan & \mc{0.18}{34} & \mc{60}{37} & \nan & \nan & \nan\\
             \makecell[lt]{RFdiffusion\\(our measurement)} & \nan & \nan & \mc{14.1}{95} & \mc{1.56}{95} & \mc{370}{95} & \nan & \nan & \nan\\
             \makecell[lt]{RFdiffusion\\(published)} & \nan & \nan & \mc{30}{95} & \mc{1400}{95} & \mc{328}{95} & \nan & \nan & \nan\\
             \bottomrule
        \end{tabular}
        \titledcaption{Experimental binding success rates and affinities of AlphaProteo model variants and RFdiffusion.}{
            Percentage of tested designs with observable binding for different AlphaProteo variants (\autoref{sec:in_silico_benchmark}) and RFdiffusion.
            Number of tested designs is in parentheses.
            Empty fields (--) are method/target combinations that were not tested.
            For success rate, AlphaProteo and "RFdiffusion (our measurement)" values come from yeast display assays performed by us, while "RFdiffusion (published)" shows BLI screening results from \citep{Watson2023-fy} (\autoref{sec:comparison_to_other_methods}).
            For AlphaProteo v1 on SC2RBD, a subset of 47 designs were made without hotspot conditioning.
            Their results are indicated separately with a dagger ($\dagger$).
            All other designs used hotspot conditioning. For {\KD} values, AlphaProteo and "RFdiffusion (our measurement)" values come from HTRF assays performed by us, while "RFdiffusion (published)" shows BLI titration results from~\citet{Watson2023-fy} (\autoref{sec:comparison_to_other_methods}).
        }
        \label{tab:internal_model_comparison}
    \end{threeparttable}
\end{table}

\begin{table}[H]
    \small
    \centering
    \renewcommand\arraystretch{1.5}
    \begin{tabular}{>{\raggedright\arraybackslash}p{0.2\textwidth}*{7}{>{\raggedleft\arraybackslash}p{0.08\textwidth}}}
        \toprule
         & BHRF1 & SC2RBD & IL-7RA & PD-L1 & TrkA & IL-17A & VEGF-A\\
        \midrule 
        Yeast surface display hits selected for \emph{E.~coli} expression & 2 & 18 & 11 & 21 & 10 & 9 & 31\\
        Successfully subcloned and expressed & 2 & 17 & 11 & 20 & 5 & 9 & 31\\
        Tested by HTRF & 2 & 17 & 11 & 18 & 5 & 7 & 31\\
        Observed HTRF binding & 2 & 14 & 11 & 16 & 5 & 3 & 20\\
        \bottomrule
    \end{tabular}
    \titledcaption{Number of yeast hits successfully expressed in \emph{E. coli} and tested for HTRF binding.}{}
    \label{tab:successful_yeast_hits}
\end{table}

\begin{table}[H]
    \small
    \centering
    \renewcommand\arraystretch{1.5}
    \begin{tabular}{>{\raggedright\arraybackslash}p{1.5cm}*{2}{>{\raggedright\arraybackslash}p{4cm}}*{2}{>{\centering\arraybackslash}p{2cm}}}
        \toprule
        Replicate & Variant & Binder & \EC ($\mu$M)& $\text{SE}(\text{EC}_{50})$\\
        \midrule
        A & ancestral            & GDM\_SC2RBD\_104 & 0.308 & 0.106 \\
         & ancestral            & GDM\_SC2RBD\_11  & 0.273 & 0.057 \\
         & ancestral            & GDM\_SC2RBD\_27  & 0.171 & 0.029 \\
         & ancestral            & GDM\_SC2RBD\_50  & 0.089 & 0.011 \\
        & BA.1                & GDM\_SC2RBD\_104 & 0.203 & 0.062 \\
         & BA.1                 & GDM\_SC2RBD\_11  & 3.053 & 4.702 \\
         & BA.1                 & GDM\_SC2RBD\_27  & 0.167 & 0.057 \\
         & XBB.1.5              & GDM\_SC2RBD\_104 & 1.958 & 1.186 \\
         \midrule
        B & ancestral            & GDM\_SC2RBD\_104 & 0.317 & 0.085 \\
         & ancestral            & GDM\_SC2RBD\_11  & 0.324 & 0.078 \\
         & ancestral            & GDM\_SC2RBD\_27  & 0.144 & 0.023 \\
         & ancestral            & GDM\_SC2RBD\_50  & 0.090 & 0.011 \\
         & BA.1                & GDM\_SC2RBD\_104 & 0.405 & 0.134 \\
         & BA.1                 & GDM\_SC2RBD\_27  & 0.382 & 0.096 \\
         & XBB.1.5              & GDM\_SC2RBD\_104 & 2.544 & 1.309 \\
         & JN.1                 & GDM\_SC2RBD\_11  & 1.019 & 0.936 \\
        \bottomrule
    \end{tabular}
    \titledcaption{\EC concentrations measured in live virus inhibition assays for SARS-CoV-2.
}{}
    \label{tab:sarscov2_inhibition}
\end{table}

\begin{table}[H]
    \small
    \centering
    \renewcommand\arraystretch{1.5}
    \begin{tabular}{>{\raggedright\arraybackslash}p{7cm}*{4}{>{\centering\arraybackslash}p{1.8cm}}}
        \toprule
        & SC2RBD\_104 & SC2RBD\_50 & SC2RBD\_11 & SC2RBD\_129\\
        \midrule
        & \multicolumn{4}{c}{\textbf{Data collection}}\\
        Microscope, operating voltage & \multicolumn{4}{c}{Titan Krios G2, 300 keV} \\
        Detector & \multicolumn{4}{c}{Falcon 4i}\\
        Automation software & \multicolumn{4}{c}{EPU}\\
        Energy filter (slit width) & \multicolumn{2}{c}{None} & \multicolumn{2}{c}{Selectris (10 eV)}\\
        Magnification (nominal) & \multicolumn{2}{c}{75,000} & \multicolumn{2}{c}{130,000}\\
        Pixel size ({\AA}) & \multicolumn{2}{c}{1.08} & \multicolumn{2}{c}{0.95}\\
        Underfocus range (nominal, $\mu$m) & \multicolumn{2}{c}{1.5 - 3.5} & \multicolumn{2}{c}{1.5 - 3.3}\\
        Number of EER frames per movie & \multicolumn{4}{c}{1,674}\\
        Total electron fluence ($e/\text{\AA}^2$) & \multicolumn{2}{c}{32.2} & \multicolumn{2}{c}{41}\\
        Total number of micrograph movies acquired & 4,500 & 8,342 & 6,728 & 8,482\\
        \midrule
        & \multicolumn{4}{c}{\textbf{Reconstruction}}\\
        Software for 2D classification & \multicolumn{4}{c}{Relion-5.0beta}\\
        Software for 3D classification & \multicolumn{4}{c}{Relion-5.0beta}\\
        Software for final reconstruction & \multicolumn{4}{c}{Relion-5.0beta, with Blush regularization}\\
        Symmetry & \multicolumn{4}{c}{C1}\\
        Number of initially extracted particles & 2,217,923 & 5,790,093 & 1,937,854 & 1,872,411\\
        Number of particles used in 3D classification & 375,214 & 960,875 & 126,130 & 385,262\\
        Number of classes in 3D classification & 7 & 6 & 4 & 5\\
        Number of particles in final reconstruction & 92,321 & 265,173 & 118,794 & 206,337\\
        Global resolution (FSC 0.143, {\AA}) & 6.0 & 4.6 & 4.7 & 4.5\\
        \bottomrule
    \end{tabular}
    \titledcaption{Cryo-EM data processing.}{}
    \label{tab:cryoem_data_processing}
\end{table}

\begin{table}[H]
    \small
    \centering
    \renewcommand\arraystretch{1.2}
    \begin{tabular}{ll}
    \toprule
    \multicolumn{2}{c}{\textbf{Data collection}}\\
                                                      Space group &                         P 41 21 2 \\
                                                      Temperature &                             100 K \\
                                               Number of crystals &                                 1 \\
             Cell dimensions ((a, b, c (\AA)),  ($\alpha$, $\beta$, $\gamma$, (\textdegree)) & (87.802 87.802 185.73) (90 90 90) \\
                                                 Wavelength (\AA) &                            0.9537 \\
    \midrule
    \multicolumn{2}{c}{\textbf{Refinement}}\\
                                           Resolution range (\AA) &          87.80 - 2.56 (2.68 - 2.56) \\
                                               No. of reflections &                      24246 (2883) \\
                                   Completeness for the range (\%) &                        100 (99.8) \\
                                                       Redundancy &                              26.6 \\
                                                           $\mathrm{R_{merge}}$ &                             0.122 \\
                                                            CC1/2 &                     0.994 (0.205) \\
                                                      Mean $I/\sigma(I)$ &                         4.4 (0.3) \\
                                           Wilson B factor ($\text{\AA}^2$) &                             47.460 \\
                                           Resolution range (\AA) &                      79.50 – 2.56 \\
                                No. observations (total/test set) &                      22961 / 1110 \\
                                                 Completeness (\%) &                             94.99 \\
                                                  Rwork/Rfree (\%) &                     0.225 / 0.227 \\
                                                     No. of atoms &                                \\
                                           \quad Protein (NON-HYDROGEN) &                              2386 \\
                                                       \quad Ligand/ion &                                 0 \\
                                                           \quad Waters &                                 0 \\
                                       Average B all atoms ($\text{\AA}^2$) &                            51.701 \\
                                                R.m.s. deviations &                                \\
                                               \quad Bond lengths (\AA) &                             0.014 \\
                                                  \quad Bond angles (\textdegree) &                              2.06 \\
                                                     Ramachandran &                                \\
                                                     \quad outliers (\%) &                                 1 \\
                                                      \quad favored (\%) &                             89.63 \\
    \multicolumn{2}{c}{$\flat$ Numbers in parentheses refer to the highest-resolution shell}\\
    \bottomrule
    \end{tabular}
    \titledcaption{Crystallographic refinement statistics for the GDM\_VEGFA\_71/VEGF-A complex structure.}{}
    \label{tab:vegfa_xtal_stats}
\end{table}

\begin{sidewaystable}[t]
    \scriptsize
    \centering
    \renewcommand\arraystretch{1}
    \begin{tabular}{>{\raggedright\arraybackslash}p{3cm}*{4}{>{\raggedleft\arraybackslash}p{0.7cm}}{>{\raggedright\arraybackslash}p{16cm}}}
        \toprule
        Design & $\mu(K_D)$ (nM) & $\sigma(K_D)$ repl. & $\sigma(K_D)$ fit & Replicas & Sequence\\
        \midrule
        BINDI (control) & 16 & 0.7 & 1 & 2 & See reference in \autoref{sec:methods}\\
        GDM\_BHRF1\_70 & 8.5 & 0.8 & 0.5 & 2 & \seqsplit{MPSAFQIGLALVAAALDRALPEPYRGLALAIAAELSGLPEEELRRLVEAAEKAASADLPFEQQVGLALARIAAAVAGVGLARRAPSLPPEELLAAIREAIEEGGRIAAKALTRSGALEPVLAELP}\\
        GDM\_BHRF1\_35 & 9.1 & 0.1 & 0.7 & 2 & \seqsplit{KEEGRKLLEEAERALRLAEELLEQGRLEAAIPPLREAILLAVKAAELGLEEEALPLLDRAADLAERGAKKARERGDKKLALEFEVLAGVALIARGVALVALRNAK}\\
        GDM\_BHRF1\_72 & 11 & 0.06 & 0.8 & 2 & \seqsplit{KEKEREQKAVSLIAAAGIALAGLEFAPQPSAEELASVLELLEEAAALSTSEEDLAFLRRLAERARELLASLPDPPAELVARLEALLARLA}\\
        LCB1 (control) & 17 & 3.2 & 2 & 8 & See reference in \autoref{sec:methods}\\
        GDM\_SC2RBD\_104 & 26 & 3.3 & 1.3 & 3 & \seqsplit{MATATLTLDKTSAKPGDTITASATGSGTATIAGARVFVVLLAFDENGNQVDSASGSAAPGETATASLTVPAGCSKVKAFAGYGDPGANKGYITDWGTVEVT}\\
        GDM\_SC2RBD\_50 & 30 & 4.5 & 2 & 3 & \seqsplit{MSAVEKAIENAKKGLENAKKDGASEESIRGLKSAINLLKEYKEGVLPESLKADAEDLIKYFSAVKD}\\
        GDM\_SC2RBD\_143 & 33 & 0.2 & 1.5 & 2 & \seqsplit{EAIEEAGRRAEEIENPDVRGAASLALGAIYAQVKNGGTGGVTAAVAVAAVANGASPSLSDEELETVARFIVDALKLLGIELPSAETLREELEAVRKAMAHSMTPEELALFDRLADALLAEVAA}\\
        GDM\_SC2RBD\_11 & 53 & 1.5 & 3.8 & 3 & \seqsplit{AAEADITLGSIIQSPSGTFAVVGGTAPAGTFPAEPTEALVKFHDGTVYHTGVTPMAMTDGTQNFSTVVPAEEAEASIGKTVTVTAGGGTVVGTLKRDPNLQVINL}\\
        GDM\_SC2RBD\_129 & 65 & 7.7 & 5.3 & 3 & \seqsplit{MATATLDAPEAAPIGTTVSATITGAPEGSTIFVTIVNLDTGLPVGSGSIRAASGTVSATIEGAKPGERYLAAAGYAADGSPVGTITAAKEFTVVE}\\
        GDM\_SC2RBD\_27 & 69 & 5.9 & 6 & 3 & \seqsplit{GNRLLAQFAGEATLEVDGETVYKGEGGFGVHDLNGRGVVTTGFNLTPEQAAKVSGTGWGTAKLVADGKEIASGPTGLVYDEESNILGANLLLSPEQAAAAGKAKTGKLEVEGTVGGKAVKMVAKGGLAESGDIPLGETA}\\
        RFD\_IL7RA\_55 (control) & 14 & 0.8 & 1.4 & 5 & \seqsplit{SELQEIAKEAGKKITEATGKKVEVEAEGNKIVIKVEEADEKTREVAEIVIEMLKDAGIEAEFEEV}\\
        GDM\_IL7RA\_70 & 0.082 & 0.007 & 0.01 & 2 & \seqsplit{MTKVEEAKELVDKIMEAAKAKDLEKVNKLRTEFFELVNSLSLEEAEEVRKYADKKGEEWYKEQL}\\
        GDM\_IL7RA\_5 & 0.49 & 0.002 & 0.04 & 2 & \seqsplit{AVEPVLSKEEVGEIARIYAKEIGKDYGIELSDEEIDLAAELARELYGKSPEEAKEFLEEVYKKLSKELSKETLKIIIAAAVGALEAAELAGRLAEEYRAGVIDADELREELSKFLPDELVDRVLARAEA}\\
        GDM\_IL7RA\_83 & 0.68 & 0.09 & 0.06 & 3 & \seqsplit{KTLLELADEFHEAVENKEYDKALAILDEIRKKYPEYKEGVDEARKRVEALKP}\\
        RFD\_PDL1\_76 (control) & 1.6 & 0.4 & 0.2 & 20 & \seqsplit{MYEVVIEGEKSVAEFIKLIAEQLGAEAEVEGDKAVIRTERREDAERLAEAAKRFGAEAEVRE}\\
        GDM\_PDL1\_135 & 0.18 & 0.0006 & 0.01 & 2 & \seqsplit{SAEEKILANLEAMKAKALAAKTEEEKLFYAKALLAVAISYAIRGDYELARRAAELAVEVIKSLSKEEQKKVMDFLINIIKNITDPEDREKAIELAIAIAERLDEEVREEALKKIEELKKE}\\
        GDM\_PDL1\_142 & 0.92 & 0.09 & 0.1 & 4 & \seqsplit{SKAEAAANRMKRFLDGLKISIPELRDLIEKYGEKIVEAIKAGDKEKALKYAEELAKKIKEVLTDDPVFAENLAKFVIVYVESLLEEL}\\
        GDM\_PDL1\_138 & 1.3 & 0.02 & 0.06 & 2 & \seqsplit{LKEEALELADEVIKLAEELGWKDHVKAVEALKEAVEKSTDERFLASAKAFLEVLKEVLLEEKKA}\\
        RFD\_TrkA\_88 (control) & 370 & - & 60 & 1 & \seqsplit{SSERAAEALRRRAEEVRQEFLDALAEIDPELAERAKEILDEGVARMEASTDEEEAARIAEEVYREITEFAPPSVHPLLDRALLLELLAFAERR}\\
        GDM\_TrkA\_9 & 0.96 & 0.1 & 0.1 & 11 & \seqsplit{APAPVLVDAGANVCKVTSGGKTSYRVLAVAGFQLPPGAGAPTVTSVTVTPHNGAAAVTIENVRAGTFSENGVTYAIVLGWAEIDAATAAALTGAPATVTVTADGKTYSKDVTIVASTATFTPA}\\
        GDM\_TrkA\_12 & 1.5 & 0.1 & 0.2 & 3 & \seqsplit{LELVSTNAPQPISGSLADGTAISGESSASVWTATESGDYPVKVTATNTGSGTVYGGGIVLAQNAGSDKLQGIGIGLTAIPPGKSVSNSGTLTVTKGGLIACAGSALCAEGGSGTLTNTITVGGKEVFSQTFTC}\\
        GDM\_TrkA\_130 & 60 & 5.8 & 6.5 & 2 & \seqsplit{SIVDELKEYFEEYKHHLSKQTKEAVEKGLADLEKILADPEKATTSEAYVFAVGAGAIAYAALKAGDKEKAEKVLELLEKVADSIPRESIRDTIRNAVRWIRRELEEYA}\\
        VEGFR1 (control) & 1.4 & 0.7 & 0.3 & 4 & See reference in \autoref{sec:methods}\\
        GDM\_VEGFA\_54 & 0.48 & 0.02 & 0.04 & 2 & \seqsplit{AEKKEKIIKALELLAEAAKKLEEAAEDPSLKEALKELKEKLKEIKEKLKKGEISLEDAANQIGALGAMIIDFADGMLAMGKIDEAEEVLKLVKEAAKALIEGGGEAGRAGRSISAKIASLEKRIAAAK}\\
        GDM\_VEGFA\_79 & 0.76 & 0.03 & 0.06 & 4 & \seqsplit{SIADIIALLEGVRDAVLAGNLDEALALMKKAADAILAEEPASPEAKALIDAAIAALEAGDFDEADAKLAEASKLIEKEGGSLAAQVVVSAMLLLGVALKSNDPALIKGVANDIGQLIDILKDWAASQ}\\
        GDM\_VEGFA\_66 & 1.5 & 0.02 & 0.2 & 2 & \seqsplit{TPEKELIEEAILALALGDREGAAAKLRELGELDPENKAFFEAQASTLLKSTNEDQLDGMMAVLLSYILEKFPLAEAEAFIEALADRVLASDAPLERKAAFLSIAASLLELEGGDPALIARLRARAAELAAQAA}\\
        GDM\_VEGFA\_71 & 4.7 & 0.06 & 0.3 & 2 & \seqsplit{GPKIHEFEGSTPGVKVVAIIGGGHAVVIAEMDIPADPAKIAKAKAALEAKAKEIEARLAPVLDRVTVHVAVDTSSNPPKAILVVELGGADAERVERLALELAKDLLEFLEKLAKELNP}\\
        IL-17RA (control) & 2.1 & 0.006 & 0.2 & 2 & See reference in \autoref{sec:methods}\\
        GDM\_IL17A\_57 & 8.4 & 0.03 & 0.9 & 2 & \seqsplit{SLLNEIRKILGEIDTIDAERFAGGDADSTPYIEKLEALVAAAPDEDLLDIARYLLELLTTPMSHDTEKAIARALIAALEKLVKKLGVKSEEIEELLERIRAAIERGEGLSGEQLDELGKILNELELIHLASKS}\\
        GDM\_IL17A\_44 & 9.1 & 0.04 & 0.6 & 2 & \seqsplit{GKTVVVDPKVDEGAARAEAEKMAKDAAPDATLMGVIKVGIGSSGDSETITVTAPDGTSISVDIPVPAFHFSALWAAPGQPDRTLTVSKTVKVPGSLTLTQDGKTKTVDVDINVKITVTGTVWDL}\\
        GDM\_IL17A\_52 & 21 & 0.4 & 2.0 & 2 & \seqsplit{SDEDWEFLKISGAKAALSNLAGIANMGFQAQLDALGDLLSAASPEVKAEAFRLIDDAQAAGVDVTPAVSLAIALAAKDLAAKGIPVNKDDLKALLDAALASVDKDLADPSKTDEQKAKLKEIKAKIEALAATI}\\
        \bottomrule
    \end{tabular}
    \titledcaption{Sequences and binding affinities of top 3 binders per target and controls.}{
        \emph{$\mu(K_D)$} is the mean fitted {\KD} value over 1-18 replicates.
        \emph{$\sigma(K_D)$ repl.} is the standard deviation of the fitted {\KD} value over replicates.
        \emph{$\sigma(K_D)$ fit} is the mean over replicates of the standard deviation estimated by fitting.
    }
    \label{tab:sequences}
\end{sidewaystable}

\clearpage

\section*{Supplementary references}
\addcontentsline{toc}{section}{Supplementary references}
\noindent Some references are listed in both the main bibliography and the supplementary bibliography section.
\begingroup
\sloppy 
\printbibliography[heading=none, title=Supplementary references]
\endgroup

\end{refsection} 

\end{document}